\pdfoutput=1

\newcommand*{\ATLASLATEXPATH}{latex/}

\documentclass[cernpreprint,texlive=2016,txfonts,UKenglish]{latex/atlasdoc}

\usepackage{\ATLASLATEXPATH atlaspackage}
\usepackage{\ATLASLATEXPATH atlasbiblatex}
\usepackage{\ATLASLATEXPATH atlascontribute}
\usepackage{\ATLASLATEXPATH atlasphysics}
\usepackage{\ATLASLATEXPATH atlascover}

\graphicspath{{logos/}{figures/}}

\usepackage{mathrsfs}
\usepackage{multirow}
\usepackage{floatrow}
\AtlasTitle{Direct top-quark decay width measurement in the $\mathbf{\boldsymbol{t\bar{t}}}$ lepton+jets channel at $\sqrt{s}=8~\text{TeV}$ with the ATLAS experiment}

\author{The ATLAS Collaboration}

 \renewcommand\Authands{, }
 \renewcommand\Affilfont{\itshape\small} 
 \AtlasContributor{Tomas Dado}{Internal note editor and main analyser}
 \AtlasContributor{Kevin Kr\"oninger}{Additional support and advice}
 \AtlasContributor{Boris Lemmer}{Analysis supervision}
 \AtlasContributor{Maria Moreno Llacer}{Analysis supervision}
 \AtlasContributor{Arnulf Quadt}{Supervisor of analyser}
 \AtlasContributor{Elizaveta Shabalina}{Analysis supervision and paper editor}
 \AtlasContributor{Philipp Stolte}{Editor and main analyser}
 \AtlasContributor{Stanislav Tokar}{Supervisor of analyser}

\AtlasRefCode{TOPQ-2017-02}

\PreprintIdNumber{CERN-EP-2017-187}

\AtlasHeading{ATLAS Paper}

\AtlasJournal{Eur. Phys. J. C}

\AtlasJournalRef{Eur. Phys. J. C 78 (2018) 129}

\AtlasDOI{10.1140/epjc/s10052-018-5595-5}

\AtlasAbstract{
  This paper presents a direct measurement of the decay width of the top 
  quark using $t\bar{t}$ events in the lepton+jets final state. The data 
  sample was collected by the ATLAS detector at the LHC in proton--proton 
  collisions at a centre-of-mass energy of 8 TeV and  
  corresponds to an integrated luminosity of 20.2~fb$^{-1}$. The decay 
  width of the top quark is measured using a template fit to distributions 
  of kinematic observables associated with the hadronically and 
  semileptonically decaying top quarks. 
  The result, $\Gamma_t = 1.76 \pm 0.33~(\text{stat.})~ ^{+0.79}_{-0.68}~(\text{syst.})~\text{GeV}$ for a top-quark mass of 172.5~GeV, 
  is consistent with the prediction of the Standard Model.
}

\AtlasCoverSupportingNote{ATL-COM-PHYS-2016-1725}{https://cds.cern.ch/record/2235510}

\AtlasCoverCommentsDeadline{x September 2017}

\AtlasCoverAnalysisTeam{Tomas Dado, Kevin Kr\"oninger, Boris Lemmer, Maria Moreno Llacer, Arnulf Quadt, Elizaveta Shabalina, Philipp Stolte, Stanislav Tokar}

\AtlasCoverEdBoardMember{Marcel Vos (chair)}
\AtlasCoverEdBoardMember{Jiri Kvita}
\AtlasCoverEdBoardMember{Mike Strauss}

 \AtlasCoverEgroupEditors{atlas-topq-2017-02-editors@cern.ch}

 \AtlasCoverEgroupEdBoard{ atlas-topq-2017-02-editorial-board@cern.ch}

\hypersetup{pdftitle={ATLAS document},pdfauthor={The ATLAS Collaboration}}

\begin{document}

\maketitle

\tableofcontents

\clearpage

\section{Introduction}
\label{sec:intro}

The top quark is the heaviest particle in the Standard Model (SM) of elementary particle physics, discovered 
more than 20 years ago in 1995~\cite{CDF, D0}. Due to its large mass of around 
173~GeV~\cite{Aaboud:2016igd, massCMS, massTevatron}, the lifetime of the top quark is extremely short. Hence, 
its decay width is the largest of all SM fermions. 
A~next-to-leading-order (NLO) calculation predicts a decay width of $\Gamma_t = 1.33$~GeV for 
a top-quark mass ($m_t$) of 172.5~GeV~\cite{Chetyrkin:1999ju,Czarnecki:1998qc}. 
Variations of the parameters entering the NLO calculation, the 
$W$-boson mass, the strong coupling 
constant $\alpha_\text{S}$, the Fermi coupling constant $G_\text{F}$ and the 
Cabibbo--Kobayashi--Maskawa (CKM) matrix element $V_{tb}$, within experimental 
uncertainties~\cite{Olive:2016xmw} yield an uncertainty of 6\%.  
The recent next-to-next-to-leading-order (NNLO) calculation predicts 
$\Gamma_t = 1.322$~GeV for $m_t=172.5$~GeV and $\alpha_\text{S} = 0.1181$~\cite{Gao:2012ja}. 

A deviation from the SM prediction could indicate non-SM decay channels of the top quark or non-SM top-quark 
couplings, as predicted by many beyond-the-Standard-Model (BSM) theories. The top-quark decay width can be 
modified by direct top-quark decays into e.g.~a charged Higgs boson~\cite{Bernreuther:2008ju,Roy:2004az} or via 
flavour-changing neutral currents~\cite{Larios:2006pb, PhysRevD.55.7241} or by non-SM radiative corrections~\cite{Herrero:2001kj}. 
Furthermore, some vector-like quark models~\cite{Alwall2007} modify the $|V_{tb}|$ CKM matrix element and 
thus $\Gamma_t$.  
Precise measurements of $\Gamma_t$ can consequently restrict the parameter space of many 
BSM models.

Extractions of $\Gamma_t$ from the $t \rightarrow Wb$ branching ratio $\mathcal{B}$ and the single-top $t$-channel 
cross-section, such as those of Refs.~\cite{PhysRevD.85.091104,Khachatryan:2014nda}, have reached a precision of 
0.14~GeV, but depend on the assumption that $\sum_q \mathcal{B} (t\rightarrow Wq) = 1$ with $q = d,s,b$, and use 
theoretical SM predictions for $\Gamma (t \rightarrow Wb)$ and the single-top $t$-channel cross-section.
Some BSM models, vector-like quark models for example~\cite{Alwall2007}, predict a sizeable deviation from the assumptions used in indirect measurements. This provides a 
motivation to perform a direct measurement of $\Gamma_t$. However, such a measurement is not yet sensitive to alternative BSM models with the current precision. 
A direct measurement of $\Gamma_t$, based on the analysis of the top-quark invariant mass
distribution was performed at the 
Tevatron by the CDF Collaboration~\cite{Aaltonen:2013kna}. 
A bound on the decay width of $1.10 < \Gamma_t < 4.05$ GeV for $m_t = 172.5$~GeV was set at 68\% confidence level.  
Direct measurements are limited by the experimental resolution of the top-quark mass spectrum, and so far are significantly less precise than indirect measurements, but avoid model-dependent assumptions.

This analysis is based on ATLAS data recorded at a centre-of-mass energy of $\sqrt{s} = 8$~TeV in 2012 in LHC proton--proton collisions. 
The top-quark decay width is extracted using $t\bar{t}$ events in the lepton+jets channel with $t\rightarrow Wb$, where 
one $W$ boson from the two top quarks decays hadronically into a pair of quarks and the 
other one decays leptonically into a charged lepton and a neutrino (the corresponding top quarks are referred to as hadronically and semileptonically decaying, respectively). 
Thus, the final state consists of four jets, two of which are $b$-jets, 
one charged electron or muon and missing transverse momentum (\met) due to the undetected neutrino. Additional jets can originate from initial- or final-state radiation (ISR or FSR).
Selected events include $W$-boson decays into a $\tau$ lepton if the $\tau$ decays leptonically.

The measurement is performed using two observables sensitive to $\Gamma_t$: \mlb, which is the 
reconstructed invariant mass of the system formed by the $b$-jet and the charged lepton $\ell$ from the 
semileptonic top-quark decay, and \bsdR, defined as the angular distance\footnote{ATLAS uses a right-handed coordinate system with its origin at the nominal interaction point (IP) in the centre of the detector and the $z$-axis along the beam pipe. The $x$-axis points from the IP to the centre of the LHC ring, and the $y$-axis points upwards. Cylindrical coordinates $(r,\phi)$ are used in the transverse plane, $\phi$ being the azimuthal angle around the $z$-axis. The pseudorapidity is defined in terms of the polar angle $\theta$ as $\eta = -\ln \tan(\theta/2)$. Angular distance is measured in units of $\Delta R \equiv \sqrt{(\Delta\eta)^{2} + (\Delta\phi)^{2}}$.} between the $b$-jet $j_b$ associated with the hadronic top-quark decay and the closest light jet $j_l$ from the hadronically decaying $W$ boson. This approach exploits the kinematic information from both the hadronically and semileptonically decaying top quarks. 
A~template method is used to measure the top-quark decay width. Templates for 
the two observables are built for all contributing SM processes. 
Distributions for multijet production 
are formed using a data-driven method. Templates for 
the other SM processes, including top-quark pair production and electroweak single-top 
production, are generated using Monte Carlo (MC) simulations. 
Templates for different top-quark decay width values are constructed by 
reweighting MC events. These templates are used in a binned likelihood fit to data to extract $\Gamma_t$. 

The ATLAS detector is described in the next section. Section~\ref{sec:samples} introduces MC 
simulated samples and the dataset, followed by a description of the event selection and 
reconstruction in Section~\ref{sec:selection}. The template fit is described in 
Section~\ref{sec:fitting}, the systematic uncertainties are estimated in Section~\ref{sec:syst}. 
Section~\ref{sec:result} presents the results of the measurement and Section~\ref{sec:conclusion} 
gives the conclusions. 

\section{ATLAS detector}
\label{sec:detector}

The ATLAS experiment~\cite{PERF-2007-01} at the LHC is a multi-purpose particle detector
with a forward-backward symmetric cylindrical geometry and a near $4\pi$ coverage in 
solid angle. 
It consists of an inner tracking detector surrounded by a thin superconducting solenoid
providing a \SI{2}{\tesla} axial magnetic field, electromagnetic and hadron calorimeters, and a muon spectrometer.
The inner tracking detector covers the pseudorapidity range $|\eta| < 2.5$.
It consists of silicon pixel, silicon microstrip, and transition radiation tracking detectors.
Lead/liquid-argon (LAr) sampling calorimeters provide electromagnetic (EM) energy measurements
with high granularity.
A hadron (steel/scintillator-tile) calorimeter covers the central pseudorapidity range ($|\eta| < 1.7$).
The endcap and forward regions are instrumented with LAr calorimeters
for both the EM and hadronic energy measurements up to $|\eta| = 4.9$.
The muon spectrometer surrounds the calorimeters and features
three large air-core toroid superconducting magnets with eight coils each.
The field integral of the toroids ranges between 2.0 and 6.0~Tm across most of the detector.
It includes a system of precision tracking chambers and fast detectors for triggering.
A three-level trigger system is used to select events.
The first-level trigger is implemented in hardware and uses a subset of the detector information
to reduce the accepted rate to at most \SI{75}{\kilo\hertz}.
This is followed by two software-based trigger levels that
together reduce the accepted event rate to \SI{400}{\hertz} on average.

\section{Data and simulated event samples}
\label{sec:samples}

The decay width of the top quark is measured using data which correspond to an integrated luminosity of $20.2~\ifb$~\cite{lumi_8TeV}. 
Single-lepton triggers for electrons and muons under stable beam conditions were used. For each lepton type, two single-lepton triggers with the transverse momentum, $\pt$, thresholds of 24 (24) and 60 (36)~GeV for electrons (muons) were used. The two triggers with the lower \pt thresholds imposed additional isolation requirements on the lepton to keep the trigger rate low. No isolation requirement was used by the higher \pt threshold triggers.

The nominal signal \ttbar MC sample was generated assuming a top-quark mass of $m_t = 172.5$ GeV using 
the \POWHEGBOX (v1) event generator~\cite{powheg,powbox1,powbox2}, referred to in the following as \POWHEG, providing NLO QCD 
matrix-element (ME)
calculations~\cite{Frixione:2007nw}. The $h_{\text{damp}}$ parameter that regulates the high-\pt radiation 
in \POWHEG was set to $m_t$. The CT10 parton distribution function (PDF) set~\cite{ct101} 
was used. The event generator was interfaced with \PYTHIA 6.425~\cite{PythiaManual} for parton showering (PS), hadronisation and underlying event modelling, using the Perugia 2011C set of tuned parameters~\cite{Skands:2010ak} and  
the CTEQ6L1 PDF set~\cite{cteq6}.
To estimate the impact of the parton shower and hadronisation model choice, a \POWHEG{}+\PYTHIA {}6 sample is compared to a 
sample generated with \POWHEG interfaced with \HERWIG 6.520~\cite{herwig} using \JIMMY v4.31~\cite{jimmy} to 
simulate the underlying event. The latter sample is referred to as \POWHEG{}+\HERWIG in the following. 
The  $h_{\text{damp}}$ parameter was set to infinity in both samples used to assess the systematic uncertainty due to parton shower modelling. The uncertainty due to the MC event generator choice is estimated using the alternative 
MC event generator \MCatNLO~\cite{Frixione:2002ik,Frixione:2003ei} for the hard process, interfaced to \HERWIG for 
showering, hadronisation and the simulation of the underlying event which is compared to the \POWHEG{}+\HERWIG sample.
To assess the impact of the initial- and final-state radiation, samples generated 
with \POWHEG were interfaced to \PYTHIA with different settings for the event generator parameters 
regulating ISR and FSR. In these samples, the  $h_{\text{damp}}$ parameter and the factorisation 
and renormalisation scales in \POWHEG, as well as the transverse momentum scale for space-like parton 
shower evolution in \PYTHIA were varied to cover the range in additional jet multiplicity corresponding 
to the uncertainty of \ttbar production measurements in association with 
jets~\cite{Aad:2015yja,Aaboud:2016omn}.
The \ttbar~samples are normalised using the theoretical cross-section 
of $\sigma_{\ttbar} = 253^{+15}_{-16}$~pb, based on a calculation performed with the 
Top++2.0~\cite{ref:xs1,ref:xs2,ref:xs3,ref:xs4,ref:xs5,ref:xs6} program that includes NNLO corrections and 
resums next-to-next-to-leading-logarithmic-order (NNLL) soft gluon terms. PDF and scale variations, the choice of $\alpha_\text{S}$, 
and the input top-quark mass are regarded as sources of systematic uncertainty. 

Background events containing a $W$ or a $Z$ boson produced in association with jets were generated 
using the \ALPGEN 2.14~\cite{alpgen} LO event generator with up to five additional partons and the 
CTEQ6L1 PDF set~\cite{cteq6}. Parton shower and hadronisation were modelled with \PYTHIA 6.425. 
Separate samples were generated for $W/Z+b\bar{b}$, $W/Z+c\bar{c}$, $W+c$, 
and $W/Z$+light jets. A parton--jet matching scheme (``MLM matching'')~\cite{mlm} is used to 
prevent double-counting of jets generated by both the matrix-element calculation and the 
parton-shower evolution. 
The $W$+jets events are normalised using a data-driven method 
exploiting the asymmetry of $W^{\pm}$ production in $pp$ collision~\cite{WjetsCA}. 
The corrections for event generator mismodelling in the fractions 
of different flavour components ($W+b\bar{b}$, $W+c\bar{c}$, $W+c$  
and $W$+light jets) are estimated in a sample with the same lepton and \met 
selections as the signal selection, but with only two jets and no $b$-tagging 
requirement. The $b$-jet multiplicity, in conjunction
with knowledge of the $b$-tagging and mistag efficiency, is used to extract
the heavy-flavour fractions.  
The correction factors extracted from the MC simulation and used in the analysis are  
$K_{b\bar{b}} = K_{c\bar{c}} = 1.50 \pm 0.11$~(stat.+syst.), $K_c = 1.07 \pm 0.27$~(stat.+syst.) and $K_{\text{light}} = 0.80 \pm 0.04$~(stat.+syst.)~\cite{WjetsCA}.
The $Z$+jets events are normalised using the inclusive NNLO theoretical cross-section~\cite{vjetsxs}.

Diboson background samples were generated with the \SHERPA 1.4.1 event generator~\cite{Gleisberg:2008ta} with up to three 
additional partons in the LO matrix elements using the CT10 PDF set.
The samples are normalised with the NLO theoretical cross-sections~\cite{dibosonxs}.

At leading order, three single-top-quark production mechanisms, $s$-channel, $t$-channel and 
associated $Wt$ production, contribute to the single-top-quark background. These processes were 
simulated with \POWHEG~\cite{Alioli:2009je,Re:2010bp} using the CT10 PDF set. All samples were interfaced to \PYTHIA 6.425 with the 
CTEQ6L1 PDF set and the Perugia 2011C tune. Overlaps between the \ttbar and $Wt$ final 
states were removed~\cite{mcatnlo_2}. All individual single-top-quark samples are normalised using their
corresponding approximate NNLO theoretical cross-sections~\cite{singleTopNorm1,stopxs_2} based on an MSTW 2008 NNLO PDF set calculation~\cite{mstw1}.

Multijet events can pass the selection because of the misidentification of a jet or a photon as an 
electron or muon (fake lepton) or because 
of the presence of a non-prompt lepton (electron or muon), which can originate from semileptonic decays 
of heavy-flavour hadrons. 
This background, referred to as multijet background in the following, is estimated directly from data using a data-driven matrix method~\cite{ttbar_3pb}. 

The detector response~\cite{atlas_sim} was simulated using the GEANT~4 simulation toolkit~\cite{geant}. 
To estimate some systematic effects, samples 
passed through a fast simulation~\cite{ATL-PHYS-PUB-2010-013} are used. 
This simulation utilises a parameterisation of the response of the EM and hadronic calorimeters 
while a full simulation is used for the tracking systems.  
The effects of in-time and out-of-time pile-up (multiple $pp$ interactions from 
the same or neighbouring bunch-crossings) are included in these simulations. 
Events from minimum-bias interactions were simulated with 
the \PYTHIA 8.1 event generator with the MSTW 2008 LO PDF set and the A2 tune~\cite{ATL-PHYS-PUB-2012-003} 
and overlaid on signal and background processes to simulate the effect of pile-up. 
The simulated events are reweighted in order to match the distribution of the average number 
of collisions per bunch crossing in the data. MC events are processed through the same reconstruction algorithms as the data.

\section{Event reconstruction and selection}
\label{sec:selection}

\subsection{Event reconstruction}
\label{sec:objects_reco}

Electrons, muons, jets, $b$-jets and missing transverse momentum are used to select $t\bar{t}$ events in this analysis.

Electron candidates are reconstructed using energy deposits in the electromagnetic calorimeter matched to reconstructed inner-detector tracks~\cite{ElectronPerformance}.
These electron candidates are required to have $\pt > 25$~GeV and $|\eta| < 2.47$, with the transition region between the barrel and endcap detector $1.37 < |\eta| < 1.52$ excluded.
Isolation requirements are used to reduce the background from fake and non-prompt electrons.
A \pt- and $\eta$-dependent isolation requirement is placed on the sum of transverse energy deposited within a cone of size $\Delta R = 0.2$ around the calorimeter cells associated to the electron. 
This energy sum excludes cells in the cluster associated with the electron and is corrected for leakage from that cluster and for energy deposits from pile-up.
Another \pt- and $\eta$-dependent isolation requirement is made on the scalar sum of track transverse momenta around the electron within a cone of size $\Delta R = 0.3$. 
Furthermore, the longitudinal impact parameter $|z_0|$ of the electron track with respect to the selected event primary vertex\footnote{The primary vertex is required to have at least four associated tracks with $\pt > 400$~MeV. If more than one vertex fulfils this criterion, the one with the largest sum of squared transverse momenta from its associated tracks is defined to be the primary vertex.} (PV) is required to be smaller than 2~mm.

Muon candidate reconstruction is based on tracks in the muon spectrometer which are matched to inner-detector tracks~\cite{MuReco}. 
The combined muon track must satisfy $\pt > 25$~GeV and $|\eta|<2.5$ and its 
longitudinal impact parameter $z_0$ with respect to the PV is required to be smaller than 2~mm.
Muon candidates have to be separated from any jet by $\Delta R > 0.4 $ and are required 
to satisfy a \pt-dependent track-based isolation requirement. Specifically, 
the scalar sum of the transverse momenta of tracks within a cone 
of size $\Delta R = 10$~GeV$/\pt$ around the muon candidate (excluding the muon track itself) has 
to be less than 5\% of the muon transverse momentum. 

Jets are reconstructed using the anti-$k_t$ algorithm~\cite{ref:Cacciari2008},
implemented in the \textsc{FastJet} package~\cite{ref:fastjet}, with a radius parameter of $R=0.4$. 
The jet reconstruction starts from calibrated topological clusters~\cite{Aad:2016upy} which are built from energy deposits in the calorimeters. 
To correct for effects due to the non-compensating calorimeter response, dead material and out-of-cluster leakage, a local cluster 
calibration scheme~\cite{LCW1,LCW2} is applied prior to jet finding. 
Simulations of charged and neutral particles are exploited to estimate these corrections. 
The jets are calibrated by applying energy- and $|\eta|$-dependent calibration factors, derived from 
simulations, to the mean energy of the jets built from the stable particles~\cite{ATLASJetEnergyMeasurement}.
In addition, a residual calibration~\cite{ATLAS-CONF-2015-037} of the jet energy scale (JES) was performed using 
data taken in 2012. Dijet events are used to calibrate 
jets in the forward region against jets in the central region. Photon+jet as well as 
$Z$+jet events are used to calibrate central jets, and multijet events 
are used to calibrate high-\pt jets. These measurements are then combined.
Jets are accepted if they fulfil $\pt > 25$~GeV and $|\eta| < 2.5$ after energy calibration.
To reduce the contribution from jets associated with pile-up, jets 
having $\pt <50$~GeV and $|\eta|<2.4$ must satisfy a requirement~\cite{Aad:2015ina} for the jet vertex fraction,\footnote{The jet vertex faction (JVF~\cite{ATLAS-CONF-2013-083}) is the ratio of the sum of the $\pt$ of tracks 
associated with the jet and the primary vertex to the sum of $\pt$ of all tracks associated 
with the jet, while the $\pt$ of the involved tracks needs to exceed $1$~GeV. Hence, this selection ensures that at least 50\% of the sum of the $\pt$ of tracks associated with the jet belongs to tracks compatible with originating from the primary vertex.} $\text{JVF} > 0.5$. 
To prevent double-counting of electrons as jets, the closest 
jet lying $\Delta R < 0.2$ of a selected electron is discarded. 
If the nearest jet surviving the selection described above is within $ \Delta R = 0.4$
of the electron, the electron is discarded.

The purity of the selected sample is improved by tagging jets containing $b$-hadrons on the basis of their large mass and decay time.
The MV1 algorithm~\cite{Aad:2015ydr} based on multivariate techniques is utilised to identify jets 
originating from the hadronisation of a $b$-quark. 
The chosen working point corresponds to an efficiency of 70\% to correctly identify a $b$-quark jet in 
simulated $t\bar{t}$ events, with a light-jet rejection factor of around 130 and a $c$-jet rejection factor of 5. The tagging efficiencies in simulation are corrected to match the results of the calibrations based on data~\cite{ATLAS-CONF-2014-004,ATLAS-CONF-2014-046}.
 
The \met serves as a measure of the transverse momentum of the neutrino 
which originates from the leptonically decaying $W$ boson. It is calculated using 
all reconstructed and calibrated particles (electrons, muons, photons) and 
jets in the transverse plane. Contributions from 
unassociated energy depositions are also taken into account~\cite{PERF-2014-04}. 

\subsection{Event selection}

According to the signature of the \ttbar signal in the lepton+jets decay channel, events are 
required to have exactly one reconstructed electron or muon 
and at least four jets, at least one of which is required to be $b$-tagged. 
This selection includes $W$-boson decays into a $\tau$ lepton if the $\tau$ decays leptonically.  
Events are required to pass a single-electron or single-muon trigger.
If at least one of the jets having $\pt > 20$~GeV is 
identified as out-of-time activity from a previous $pp$ collision, as 
calorimeter noise or non-collision background, the event is not considered~\cite{ATLAS-CONF-2012-020}.

Events with exactly one $b$-tagged jet need to have \met > 20 GeV and $\met + m^W_{\text{T}} > 60$ GeV, 
where $m^W_{\text{T}}$ is the transverse mass of the leptonically decaying $W$ boson defined as 
$m^W_{\text{T}} = \sqrt{ 2p^{\ell}_{\text{T}} \met (1-\cos \Delta \phi (\ell,\met)) }$.  
These requirements suppress the background due to misidentified leptons. As this background becomes 
very small in high $b$-tag multiplicity regions, these requirements are not 
applied to events with 
at least two $b$-tagged jets.  
Selected events are reconstructed under the \ttbar decay hypothesis 
using a likelihood-based method described in Section~\ref{sec:eventreco}.
The logarithm of the likelihood has to satisfy  $\ln(L)> -50$ to suppress 
the combinatorial background due to wrongly reconstructed events and to decrease other 
backgrounds. This improves the sensitivity of the measurement by increasing the 
fraction of well-reconstructed \ttbar events in the selected sample.
Events satisfying all selection criteria are separated into eight mutually exclusive analysis regions.
The events are categorised according to the flavour of the selected lepton and whether they have exactly one or at least two $b$-tagged jets. They are further split
into two $|\eta|$ regions, a central region 
with all four jets associated with the \ttbar decay 
having $|\eta| \leq 1$ and a second one with at least 
one jet with $|\eta| > 1$. 
This approach takes advantage of the different sensitivity of these regions to detector resolution 
effects and pile-up contributions and different amounts of background.
The corresponding event yields are listed in Table~\ref{tab:event_yield_eta}. Figures~\ref{fig:control_plots_el_2incl} and 
~\ref{fig:control_plots_mu_2incl} show the distributions of the lepton and leading $b$-tagged jet \pt, lepton and leading $b$-tagged jet $\eta$, \met and $m^W_{\text{T}}$ for events with at least two $b$-tagged jets in the
electron and muon channels, respectively. 
Good agreement within the assigned statistical and systematic uncertainties is observed between data and the predictions from simulation.

\begin{table}[t]
\centering 
\caption{Event yields after the event selection in the (a) electron+jets and (b) muon+jets channel 
for events with exactly one or at least two $b$-tags divided into events where all four 
jets associated with the \ttbar decay have $|\eta| \leq 1$ and events where at least one 
jet has  $|\eta| > 1$. The uncertainties in the signal and background yields arising from 
normalisation uncertainties of each sample are shown. These correspond to the theory 
uncertainties as 
described in Section~\ref{sec:fitting} for the background sources except for the $W$+jets and 
the multijet background, whose uncertainties originate from the data-driven methods.}
\subfloat[Electron+jets channel.]{
\begin{tabular}{|l|r|r|r|r|}
\hline
$e$+jets & \multicolumn{2}{c|}{$|\eta| \leq 1$ region} & \multicolumn{2}{c|}{$|\eta| > 1$ region} \\\hline
Sample &  1 $b$-tag & $\geq 2$ $b$-tags &  1 $b$-tag & $\geq 2$ $b$-tags  \\
\hline
\ttbar     	&  5850	$\pm$ 380	& 6480~~~	$\pm$ 420~~~	&  29200	$\pm$ 1900	& 27600 $\pm$ 1800	\\
Single top 	&  285 	$\pm$ ~~48 	& 141~~~	$\pm$ ~~24~~~	&  1830		$\pm$ ~~310  	& 860 	$\pm$ ~~150	\\
$W$ + $bb/cc$	&  362 	$\pm$ ~~40 	& 81~~~	$\pm$ ~~~~9~~~	&  2640 	$\pm$ ~~290 	& 506 	$\pm$ ~~~~56 	\\
$W$ + $c$ 	&  174 	$\pm$ ~~47 	& 8~~~	$\pm$ ~~~~2~~~	&  1300 	$\pm$ ~~350 	& 56 	$\pm$ ~~~~15 	\\
$W$ + light	&  87 	$\pm$ ~~~~3 	& 3.7	$\pm$ ~~~~0.2	&  578	 	$\pm$ ~~~~23  	& 26 	$\pm$ ~~~~~~1 	\\
$Z$ + jets 	&  120 	$\pm$ ~~58 	& 38~~~	$\pm$ ~~18~~~	&  1190 	$\pm$ ~~570 	& 310  	$\pm$ ~~150 	\\
Diboson    	&  31 	$\pm$ ~~15 	& 4~~~	$\pm$ ~~~~2~~~	&  183 		$\pm$ ~~~~88  	& 29 	$\pm$ ~~~~14 	\\
Multijet  	&  228 	$\pm$ ~~68 	& 38~~~	$\pm$ ~~11~~~	&  2490 	$\pm$ ~~750 	& 540 	$\pm$ ~~160 	\\
\hline
Total expected  &  7140	$\pm$ 400 	& 6790 	$\pm$ 420~~~ 	&  39400 	$\pm$ 2200	& 29900 $\pm$ 1800	\\
\hline
Data       	& 6800 \hspace{9mm}  	& 7056 \hspace{12mm}			&  37823 \hspace{11mm}          	& 30644 \hspace{11mm}     		\\
\hline
\end{tabular} }
\vspace{2mm}
\subfloat[Muon+jets channel.]{
\begin{tabular}{|l|r|r|r|r|}
\hline
$\mu$+jets & \multicolumn{2}{c|}{$|\eta| \leq 1$ region} & \multicolumn{2}{c|}{$|\eta| > 1$ region} \\\hline
Sample &  1 $b$-tag & $\geq 2$ $b$-tags &  1 $b$-tag & $\geq 2$ $b$-tags  \\
\hline
\ttbar     	&  7000 $\pm$ 450	& 7640~~~	$\pm$ 490~~~	&  35900	$\pm$ 2300	& 33500 $\pm$ 2200	\\
Single top 	&  369 	$\pm$ ~~63  	& 160~~~	$\pm$ ~~27~~~	&  2110 	$\pm$ ~~360 	& 980 	$\pm$ ~~170 	\\
$W$ + $bb/cc$	&  473 	$\pm$ ~~52 	& 117~~~	$\pm$ ~~13~~~	&  3450 	$\pm$ ~~380 	& 756 	$\pm$ ~~~~83 	\\
$W$ + $c$ 	&  223 	$\pm$ ~~60 	& 5~~~	$\pm$ ~~~~1~~~	&  1540 	$\pm$ ~~420 	& 63 	$\pm$ ~~~~17 	\\
$W$ + light	&  96 	$\pm$ ~~~~4 	& 1.8	$\pm$ ~~~~0.1	&  797 		$\pm$ ~~~~32 	& 40 	$\pm$ ~~~~~~2	\\
$Z$ + jets 	&  74 	$\pm$ ~~36 	& 16~~~	$\pm$ ~~~~8~~~	&  610	 	$\pm$ ~~290 	& 159  	$\pm$ ~~~~76 	\\
Diboson    	&  37 	$\pm$ ~~18 	& 6~~~	$\pm$ ~~~~3~~~	&  198 		$\pm$ ~~~~95 	& 32 	$\pm$ ~~~~15 	\\
Multijet   &  195 	$\pm$ ~~59 	& 34~~~	$\pm$ ~~10~~~	&  1870 	$\pm$ ~~560 	& 400 	$\pm$ ~~120 	\\
\hline
Total expected  &  8470 $\pm$ 470 	& 7980	$\pm$ 490~~~	&  46400 	$\pm$ 2500  	& 36000 $\pm$ 2200	\\
\hline
Data       	&  8274	\hspace{9mm}    & 8193 \hspace{12mm}			&  46275 \hspace{11mm}	            	& 36471 \hspace{11mm}      		\\
\hline
\end{tabular} }

\label{tab:event_yield_eta}
\end{table}

\begin{figure}[!hbpt]
\begin{center} 
	\includegraphics[width=67mm]{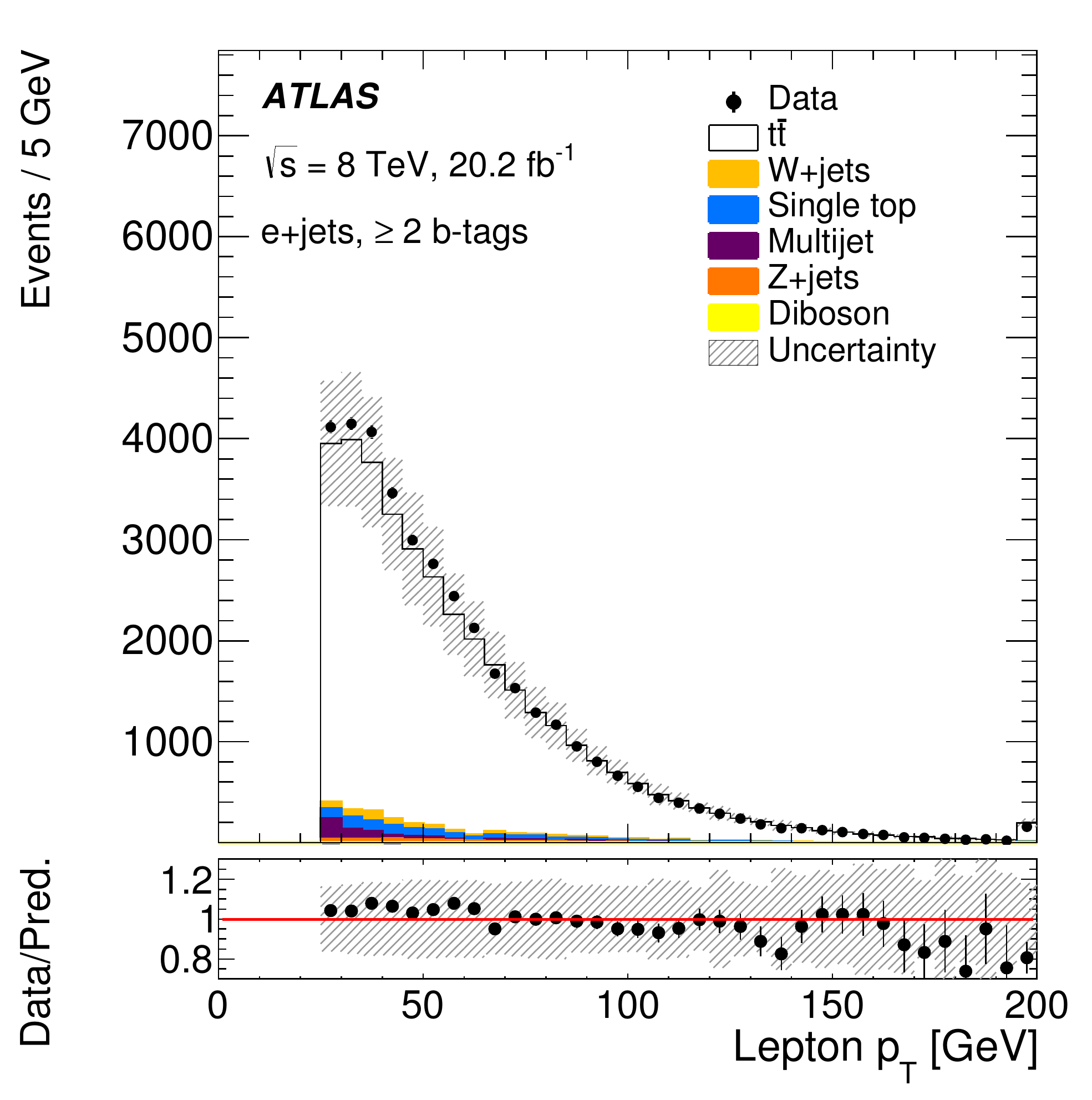}
	\includegraphics[width=67mm]{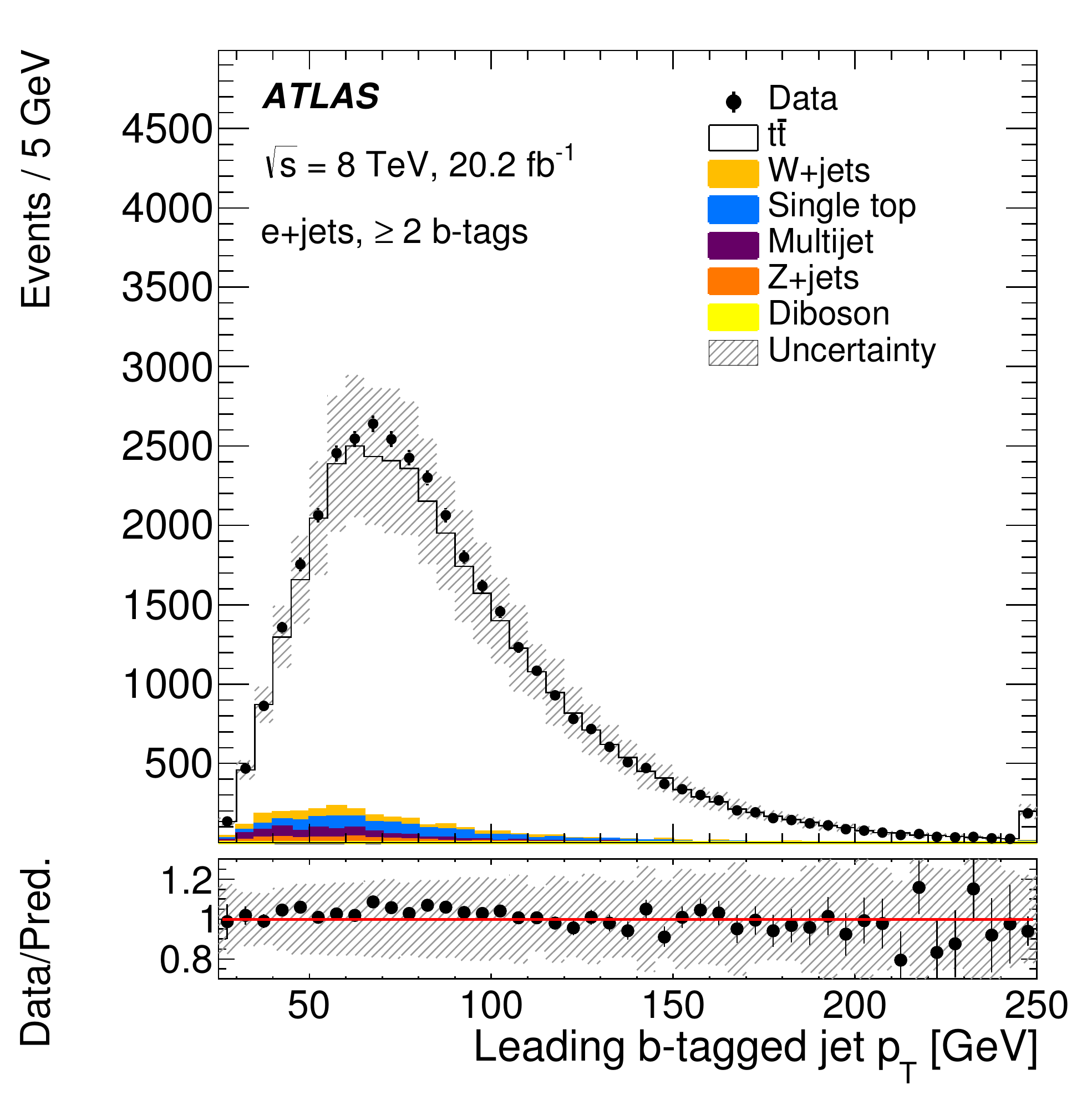}\\
	\includegraphics[width=67mm]{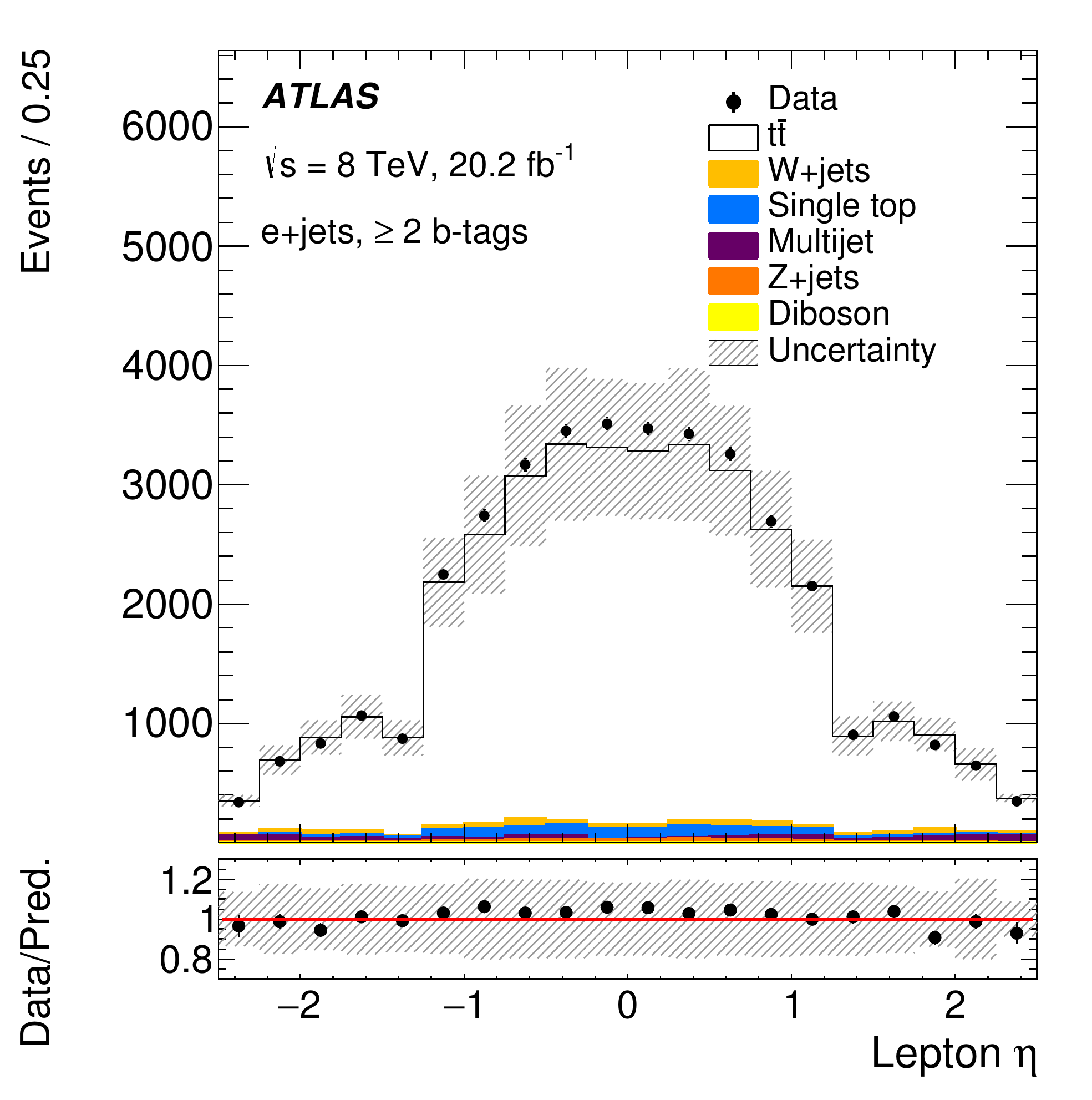}
	\includegraphics[width=67mm]{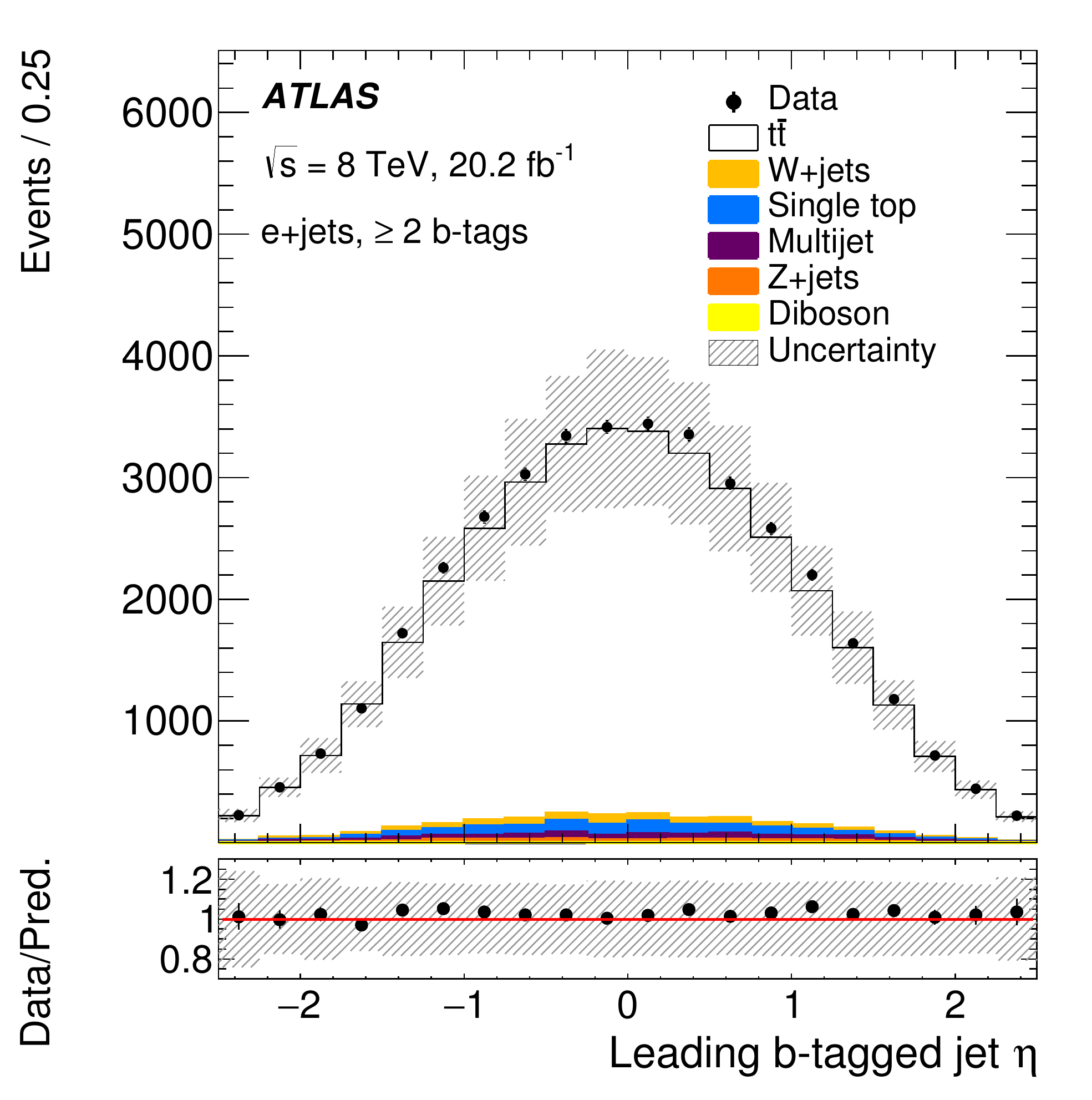}\\
	\includegraphics[width=67mm]{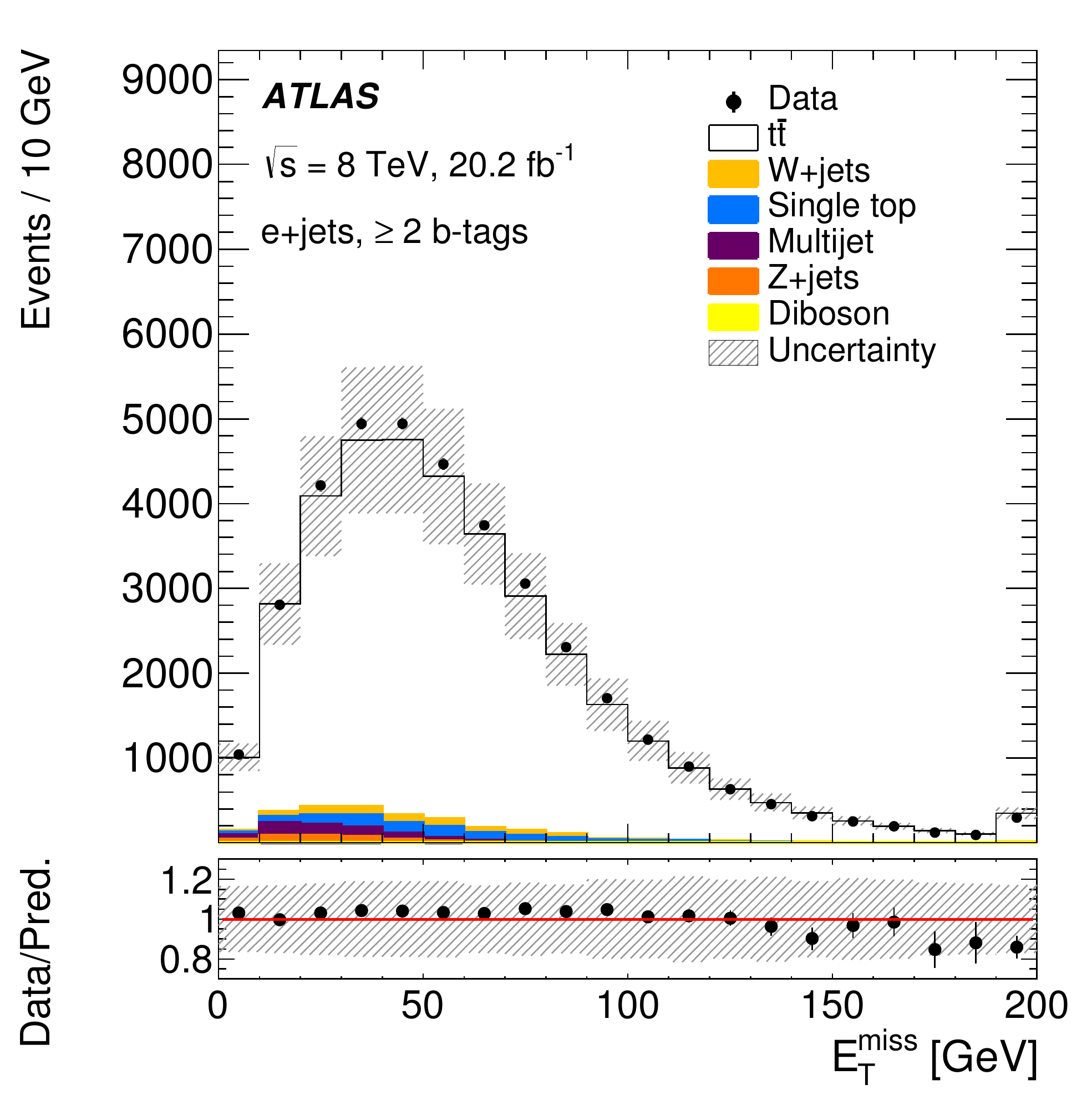}
        \includegraphics[width=67mm]{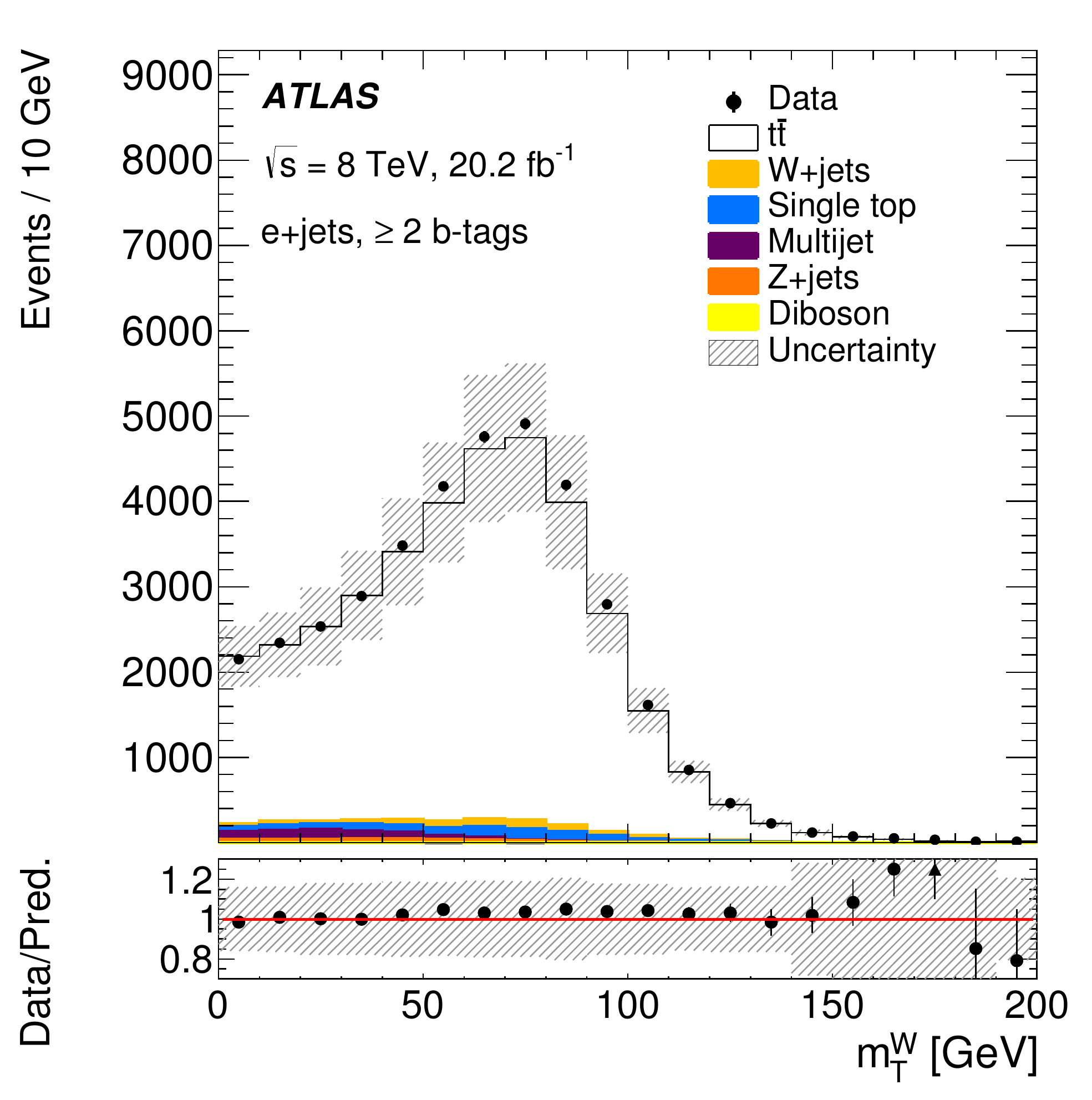}
	\caption{Distributions of the lepton and leading $b$-tagged jet \pt, lepton and leading $b$-tagged jet $\eta$, \met 
	and $m^W_{\text{T}}$ in the  
	electron+jets channel for events with at least two $b$-tagged jets after event selection. 
	The hatched bands show the normalisation 
	uncertainty in the signal and background contributions and the signal model systematic uncertainties. The first and last bins contain 
	underflow and overflow events, respectively.}
\label{fig:control_plots_el_2incl}
\end{center}
\end{figure}

\begin{figure}[!hbpt]
\begin{center} 
	\includegraphics[width=67mm]{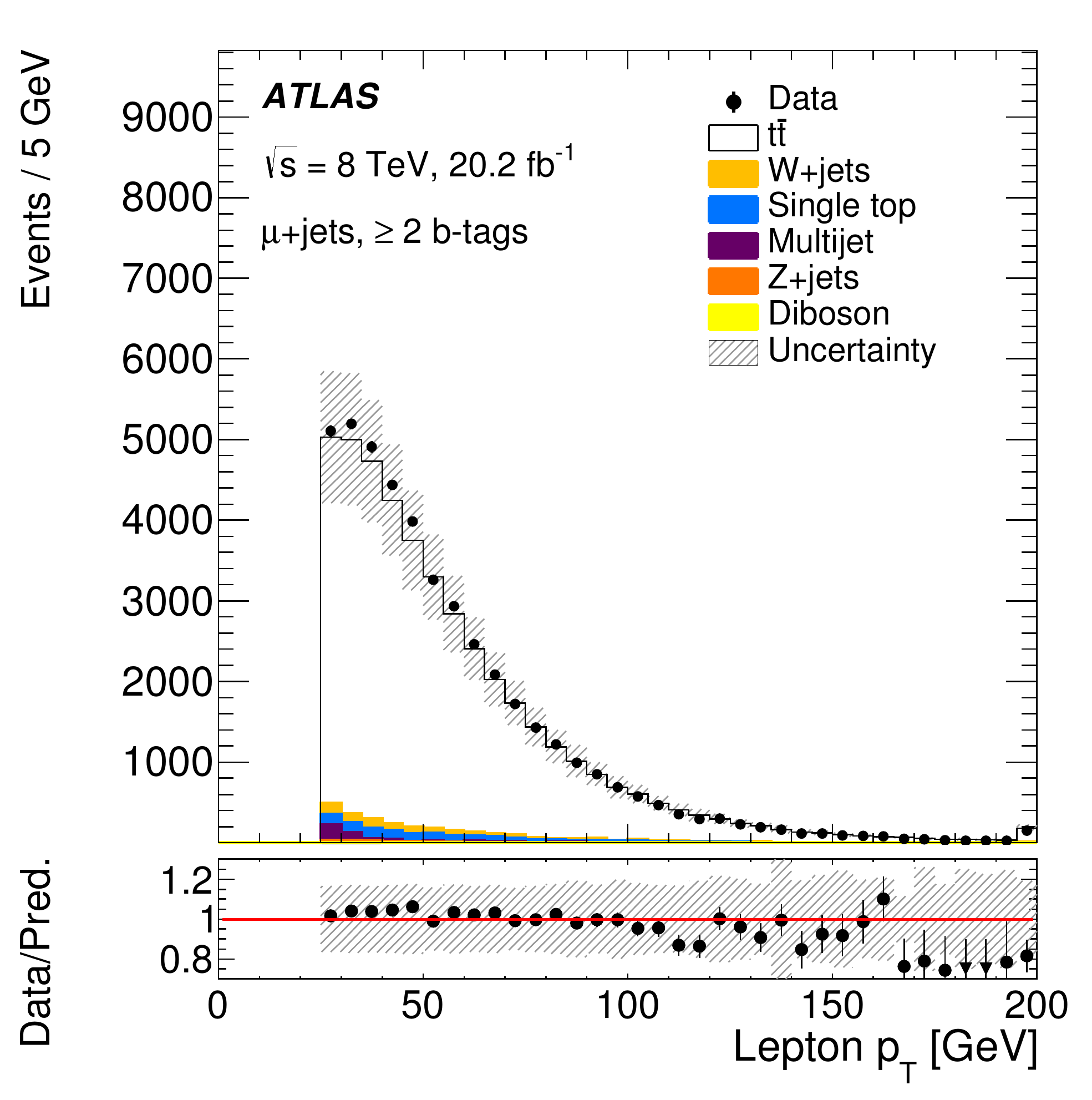}
	\includegraphics[width=67mm]{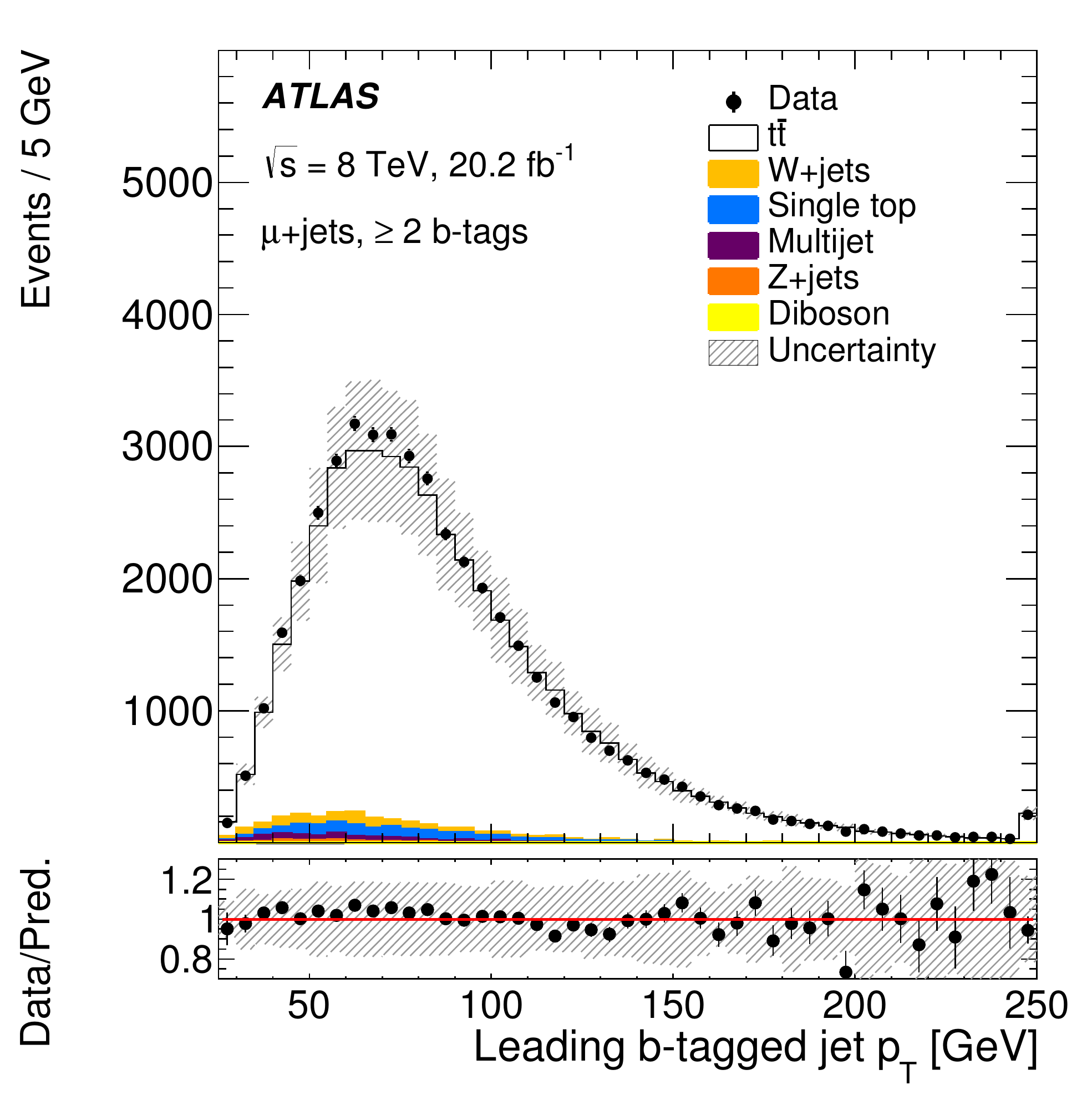}\\
	\includegraphics[width=67mm]{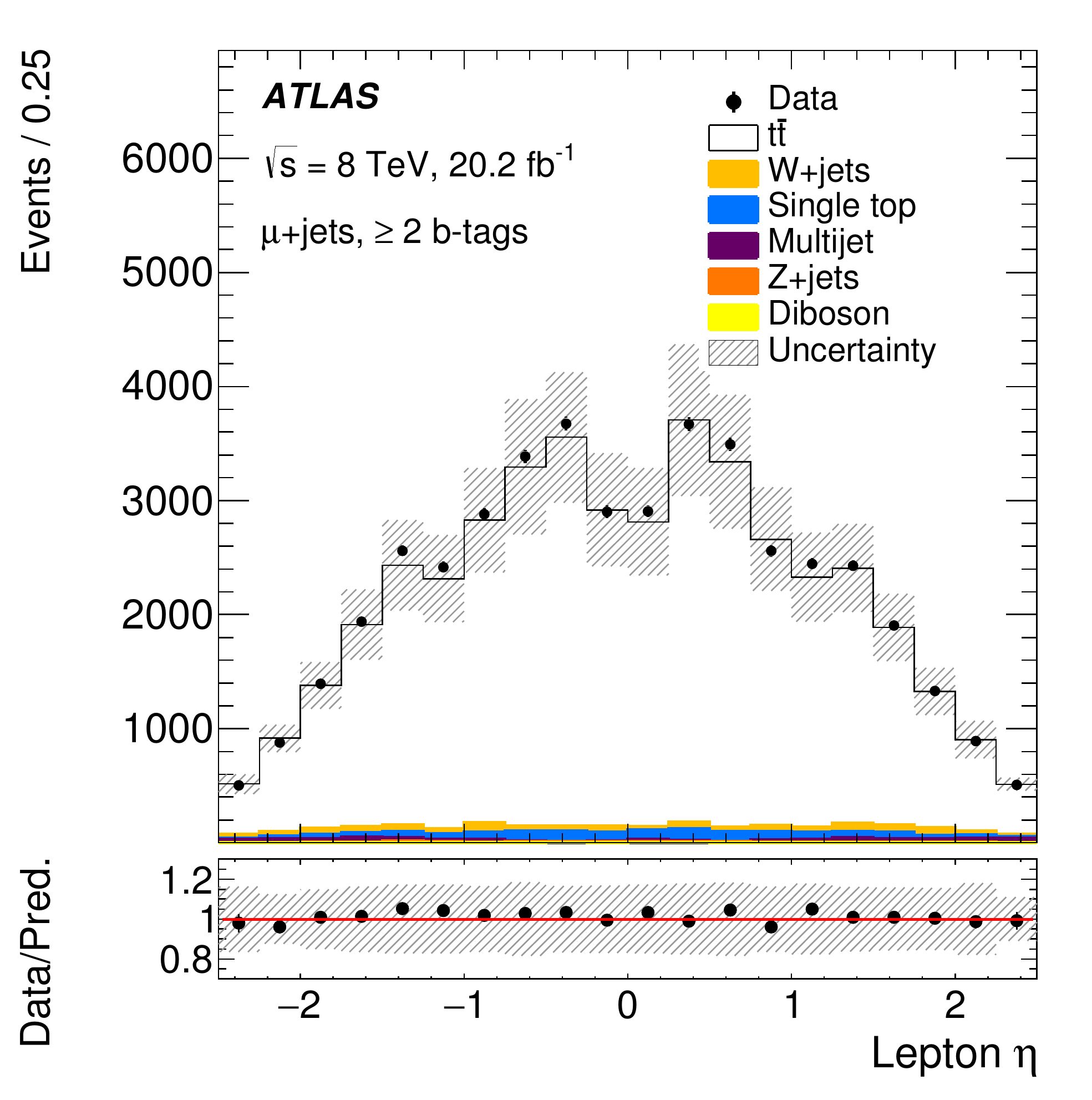}
	\includegraphics[width=67mm]{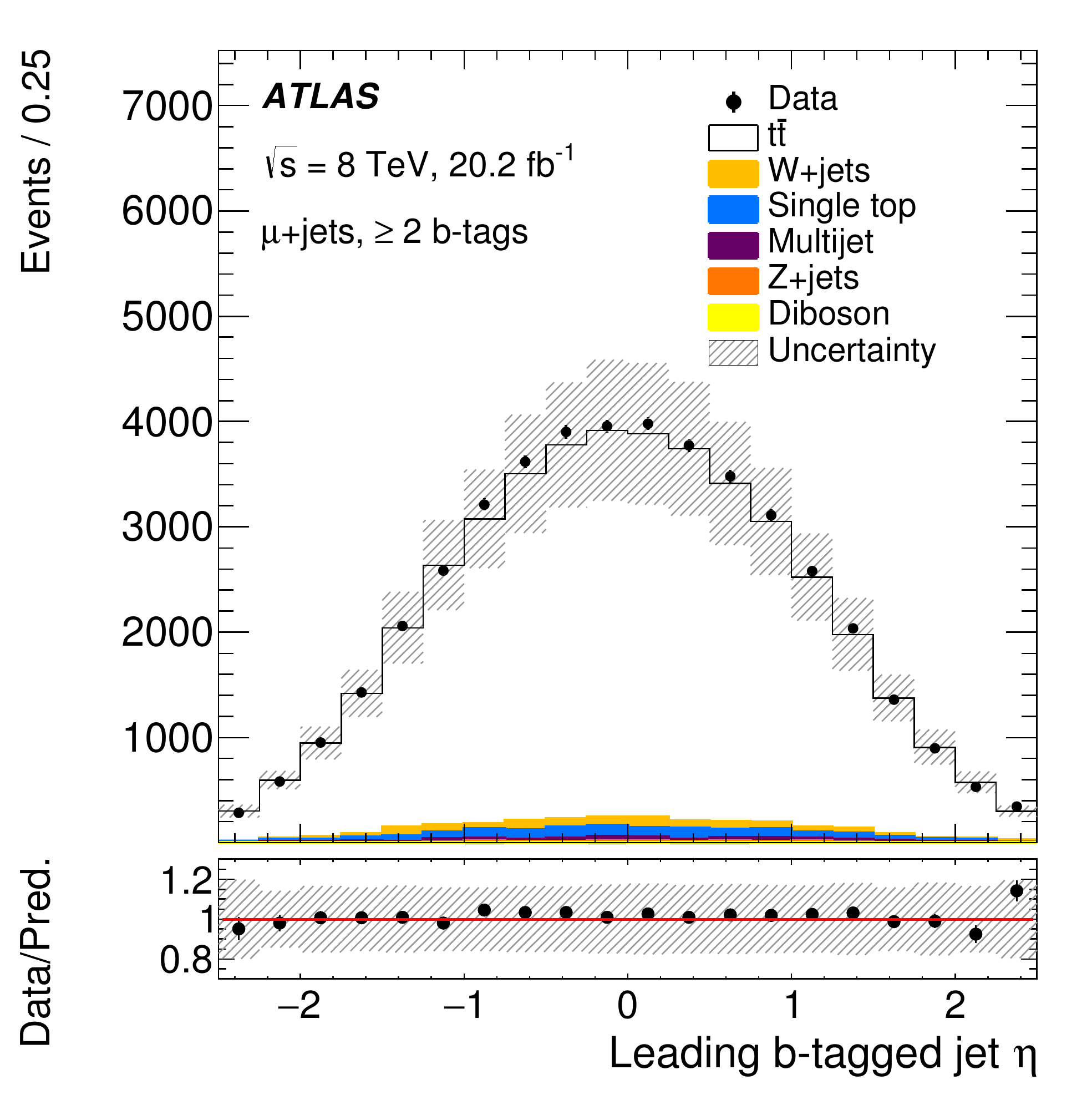}\\
	\includegraphics[width=67mm]{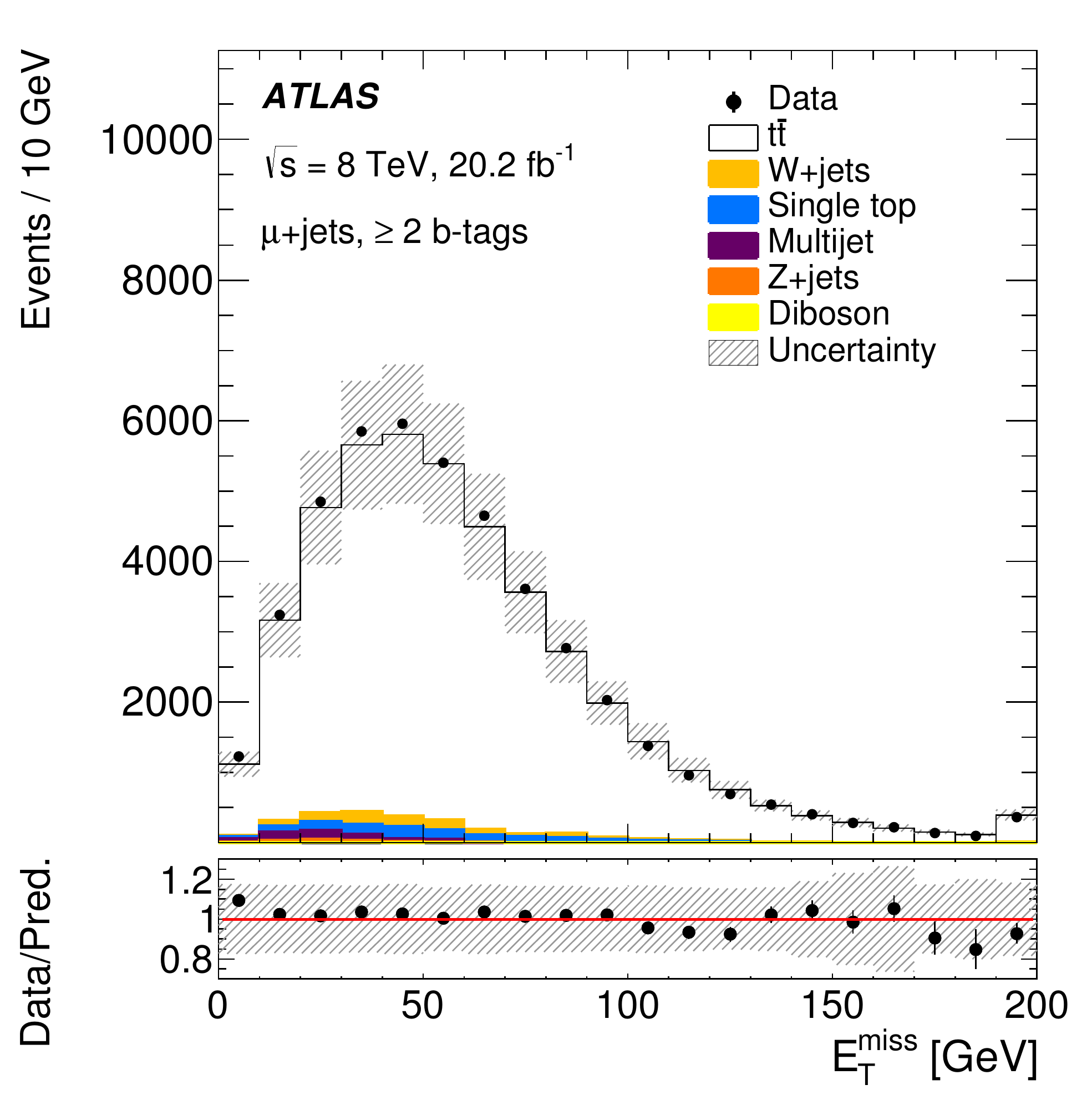}
        \includegraphics[width=67mm]{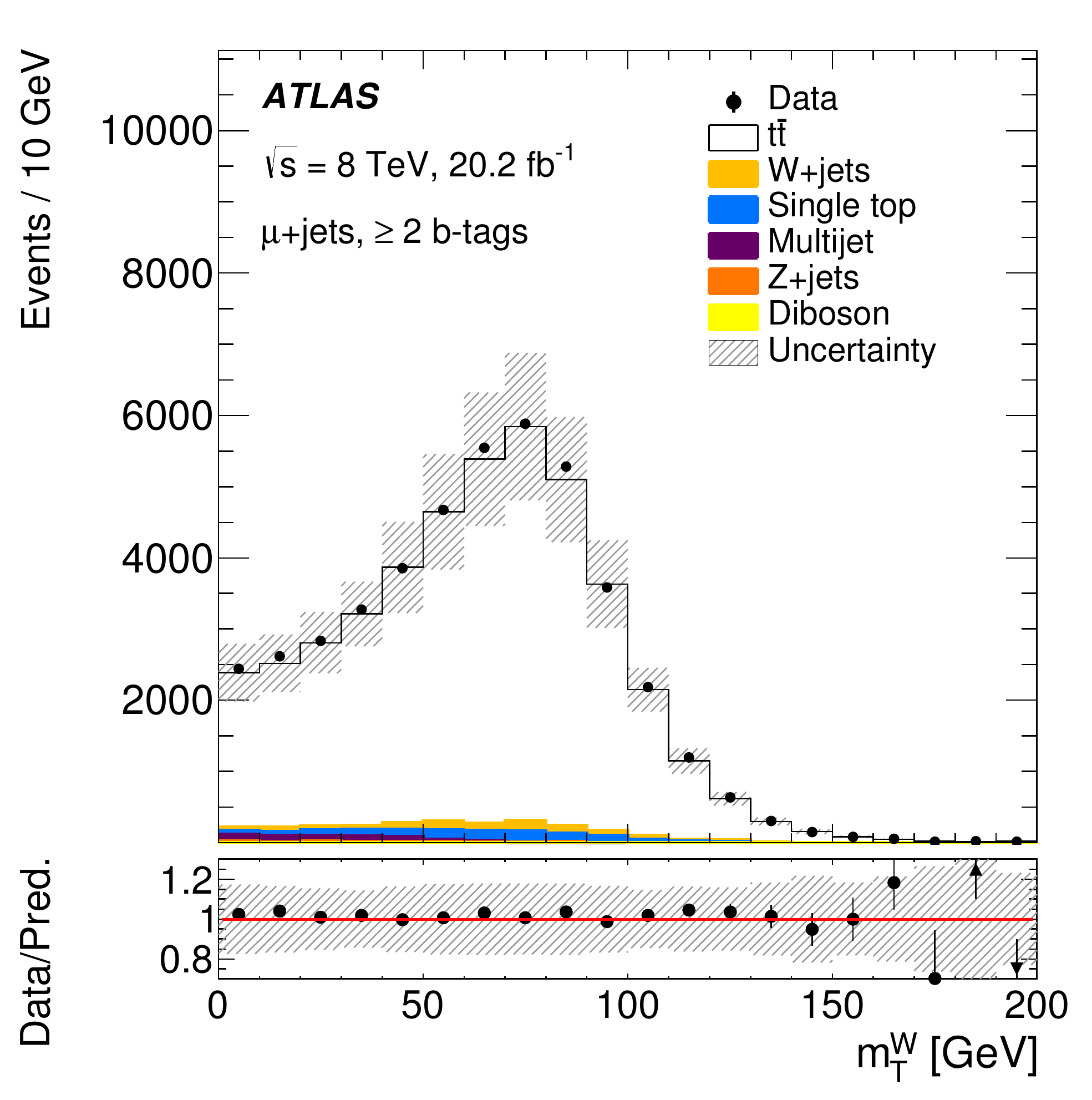}
	\caption{Distributions of the lepton and leading $b$-tagged jet \pt, lepton and leading $b$-tagged jet $\eta$, \met 
	and $m^W_{\text{T}}$ in the  
	muon+jets channel for events with at least two $b$-tagged jets after event selection. The hatched 
	bands show the normalisation
	uncertainty in the signal and background contributions and the signal model systematic uncertainties. The first and last bins contain 
	underflow and overflow events, respectively.}
\label{fig:control_plots_mu_2incl}
\end{center}
\end{figure}

\subsection{Reconstruction of the \ttbar decay}
\label{sec:eventreco}

The correct assignment of reconstructed jets to partons originating from a \ttbar decay is important for this measurement. 
This is achieved using a likelihood-based method (Kinematic Likelihood Fitter, KLFitter~\cite{klfitter}) which makes 
use of the Bayesian Analysis Toolkit~\cite{Caldwell:2008fw}.
KLFitter maps the four partons of the \ttbar decay to four reconstructed jets  
using mass constraints on the top-quark mass $m_t$ and the $W$-boson mass $m_W$. In this analysis the
four jets with the highest $p_{\text{T}}$ are used as input to KLFitter. 
A likelihood $L$ is maximised for 
all resulting 24 permutations. For each permutation the likelihood is defined as

\begin{align}
L &= \textrm{BW}(m_{q1q2q3}| m_t,\Gamma_{t})\cdot \textrm{BW}(m_{q1q2}| \mW,\Gamma_{W})\cdot \textrm{BW}(m_{q4\ell\nu}| m_t,\Gamma_{t})\cdot \textrm{BW}(m_{\ell\nu} | \mW,\Gamma_{W})  \nonumber \\
&\cdot\prod\limits_{i=1}^4 W(E_{i}^{\text{meas}}|E_{i})\cdot W(E_{\ell}^{\text{meas}}|E_{\ell})\cdot
W(E_{x}^{\text{miss}}|p_{x}^{\nu})\cdot W(E_{y}^{\text{miss}}|p_{y}^{\nu})\,.
\label{eq:KLFlikelihood}
\end{align}

The $W(E_{P}^{\text{meas}}|E_{P})$ are transfer functions, where $E_{P}^{\text{meas}}$ is the measured energy of the jet or lepton $P$, $E_{P}$ is the energy of the corresponding parton or lepton, and $p_{x}^{\nu}$ and $p_{y}^{\nu}$ are the momentum components of the neutrino $\nu$ in the transverse 
plane. These momentum components as well as the energies $E_P$ are free parameters of the likelihood maximisation. 
The component $p_{z}^{\nu}$ is initially calculated using a constraint on the $W$-boson mass $m_W^2 = (p_\nu + p_\ell)^2$ with the four-momenta $p_\nu$ and $p_\ell$~\cite{klfitter}.
Transfer functions for electrons, muons, $b$-jets, light jets (including $c$-jets) and \met are used. 
They are derived from simulated \ttbar events using \MCatNLO{}+\HERWIG~\cite{Frixione:2002ik,Frixione:2003ei,herwig}. The decay products of the \ttbar pair are 
uniquely matched to reconstructed particles to obtain a continuous function which describes the 
relative energy difference between a parton and a reconstructed jet or particle as a function of the parton energy.  
Parameterisations are derived for different $|\eta|$ regions.
The $\textrm{BW}(m_{ij(k)}| m_{t/W},\Gamma_{t/W})$ terms represent Breit--Wigner functions which stand for the 
probability distribution of the reconstructed $W$-boson or top-quark mass given the assumed mass  
$m_{t/W}$ and a decay width $\Gamma_{t/W}$. Indices $q1$--$q4$ refer to the four quarks mapped to the 
reconstructed jets.

To exploit the presence of two $b$-quarks in a \ttbar decay, kinematic information is  
complemented by $b$-tagging. To take it into account, the likelihood definition of 
Eq.~(\ref{eq:KLFlikelihood}) is extended and turned 
into an event probability which, for a given permutation $i$, is expressed as

\begin{align*}
 P_{i}= \frac{L_i\prod_j p_{i,j}}{\sum_k L_k\prod_j p_{k,j}}\,. 
\end{align*}

The $p_{i,j}$ contain the $b$-tagging efficiency or the mistag rate corresponding to the $b$-tagging working point, depending on the jet $j$ flavour assigned by KLFitter and whether it is $b$-tagged or not.
This factor is calculated for all jets $j$ and multiplied by the likelihood $L_i$. 
KLFitter calculates the latter quantity for each permutation in the event according to 
Eq.~(\ref{eq:KLFlikelihood}). 
The permutation with the largest event probability determines the jet-to-parton assignment that is used to build the observables \mlb and \bsdR.
In this analysis the mass parameters are set to $m_W=80.4~\gev$ and $m_t=172.5~\gev$ and the decay width parameters are fixed to $\Gamma_W = 2.1$~GeV and $\Gamma_t = 1.33$~GeV. 
The analysis uses KLFitter only to choose the best assignment of jets
to partons and does not exploit the fitted four-momenta for the
reconstructed particles. A variation of the $\Gamma_t$ parameter used in KLFitter 
was proven to have no impact on the reconstructed distributions and
thus the extracted measured value of $\Gamma_t$.

Figure~\ref{fig:control_plots_LL} shows distributions of the logarithm of the likelihood 
for different analysis regions. 
Fully matched \ttbar events populate the high $\ln(L)$ region. Thus, a requirement of $\ln(L)> -50$ removes a significant fraction of the combinatorial background. However, both background events and \ttbar events with partially correctly and incorrectly matched jets contribute to the full range of likelihood values. The double peak structure of the output is thus not related to a correct match of jets but caused by the migration of the events which are not matched correctly towards the higher values of the likelihood due to the fixed top quark mass requirement.
The fraction of events where all four partons are matched correctly  
increases from 13\% to 23\% ( 17\% to 31\%) after applying this requirement to events with at least one (two) $b$-tagged jet(s). 
This selection also improves the purity of the sample
by removing more background events than \ttbar signal. 
The analysis does not rely on matching correctly all four jets. 
The observable \mlb, which provides most of the sensitivity to $\Gamma_t$, 
depends solely on the correct assignment of the $b$-jet from the semileptonically 
decaying top quark for which the reconstruction efficiency is 65\% (75\%) for events with at least one (two) $b$-tagged jet(s).

\begin{figure}[!hbpt]
\begin{center} 
	\subfloat[]{\includegraphics[width=70mm]{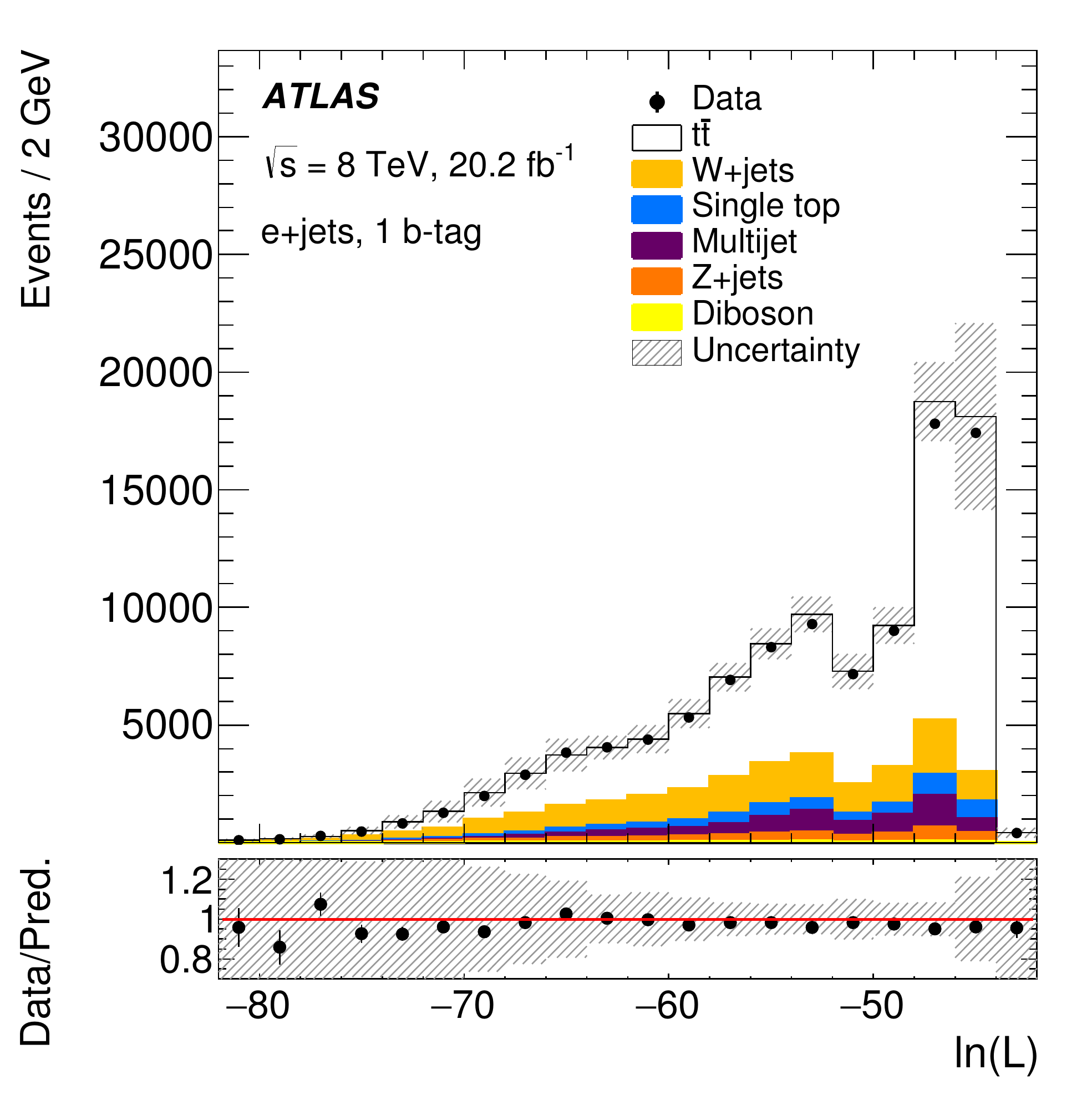} }
	\subfloat[]{\includegraphics[width=70mm]{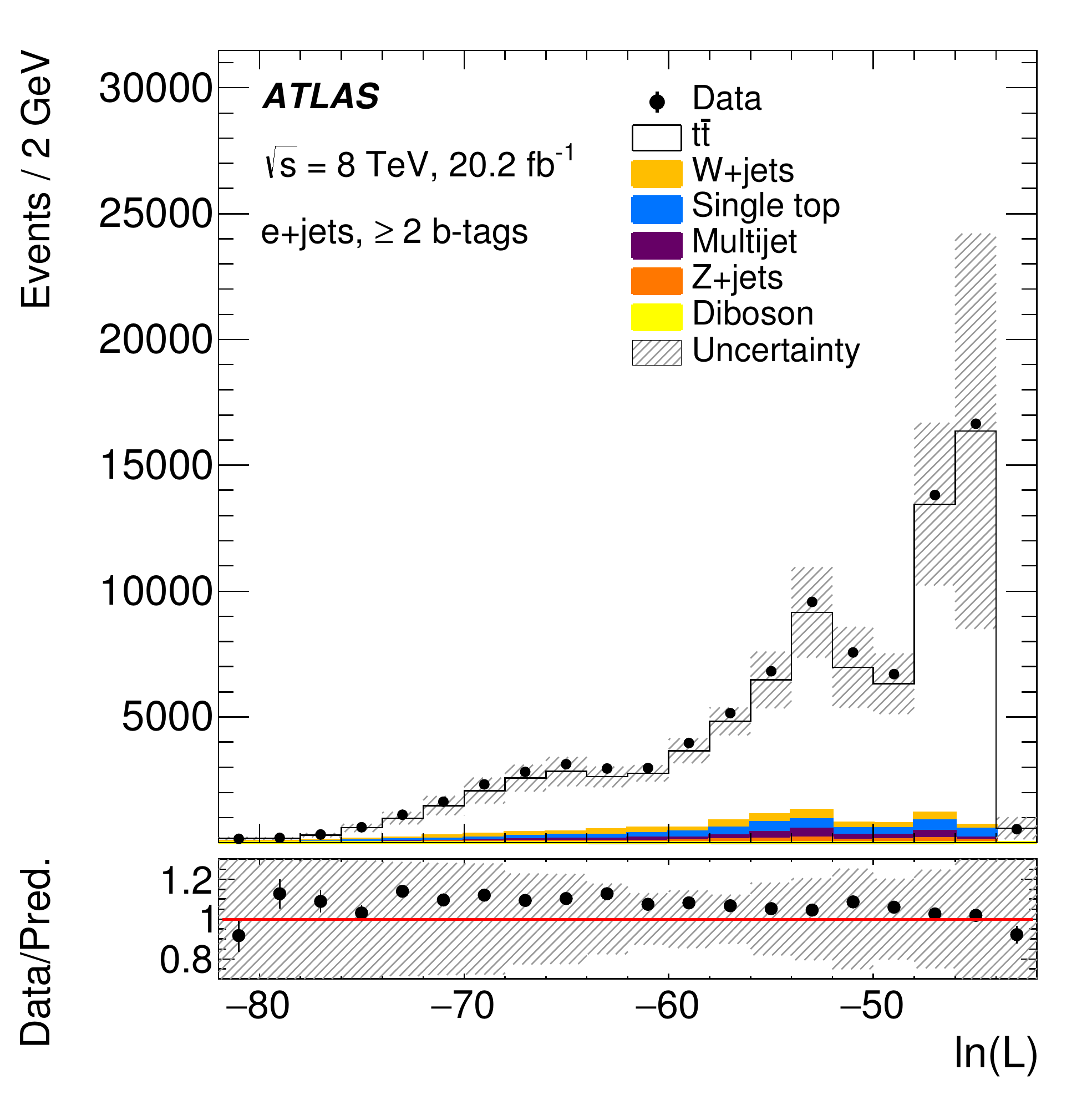} }\\
	\subfloat[]{\includegraphics[width=70mm]{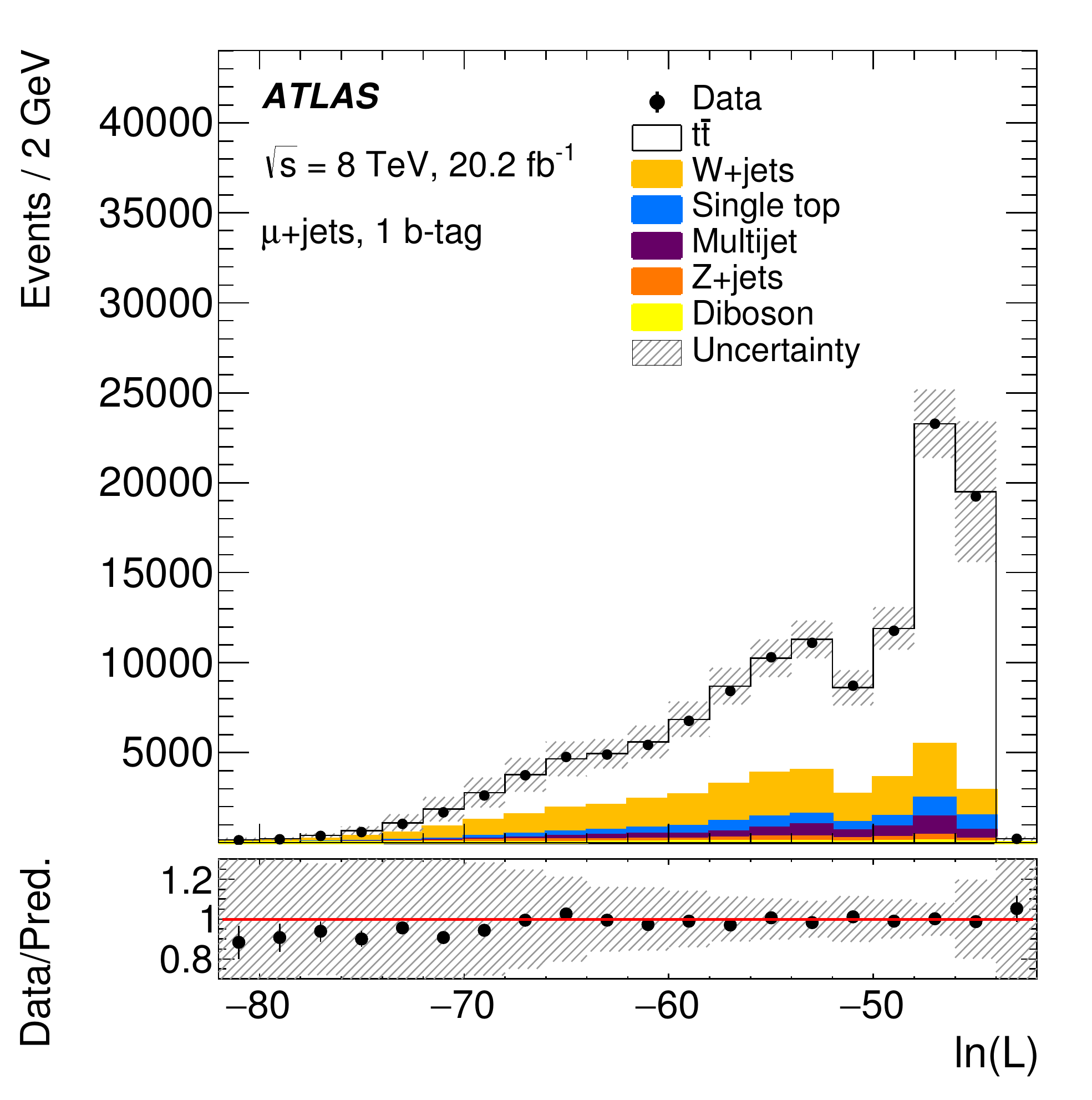} }
	\subfloat[]{\includegraphics[width=70mm]{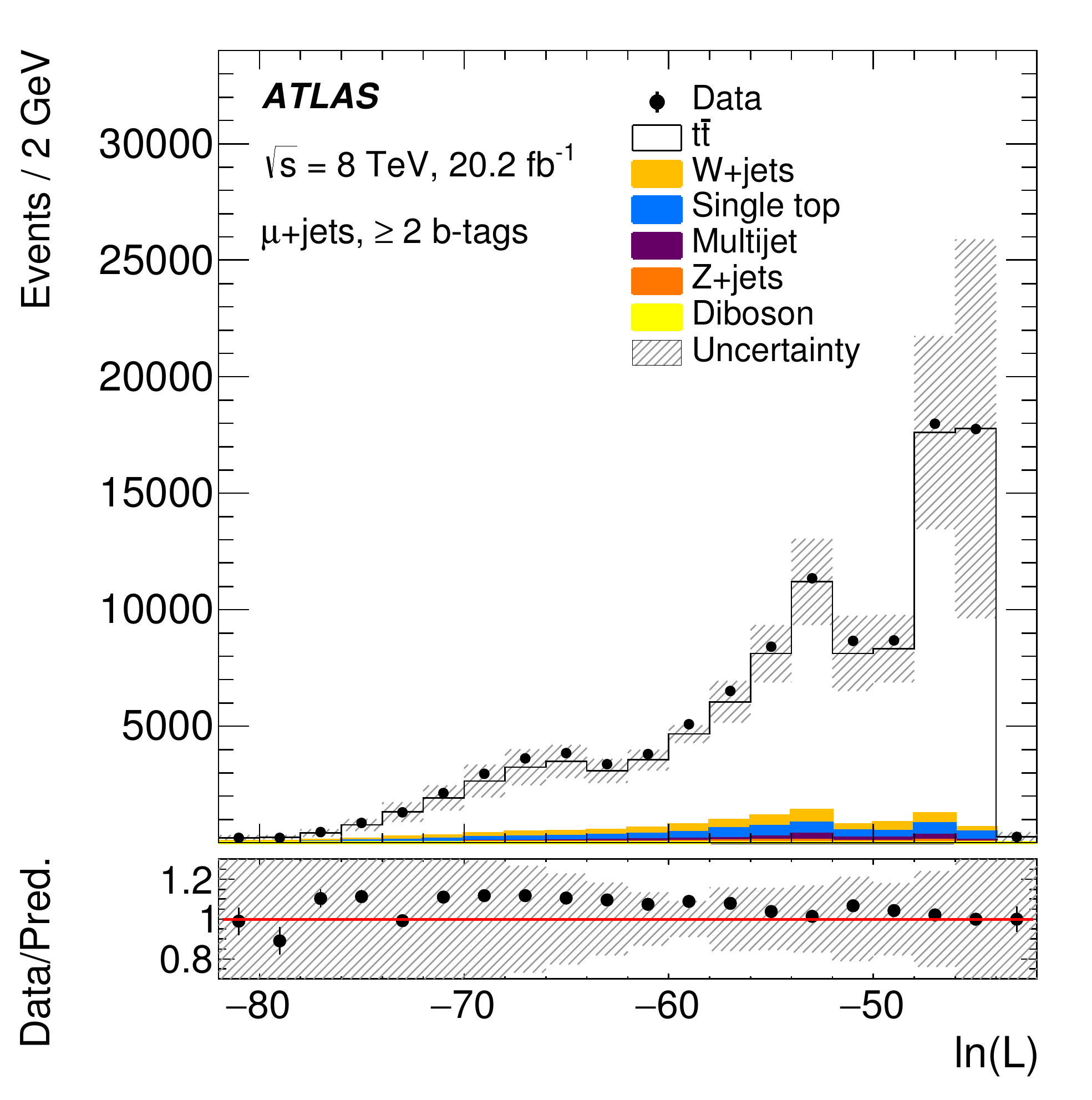} }
	\caption{Distributions of the logarithm of the likelihood obtained from the 
	event reconstruction algorithm for the selected (a,b) 
	electron+jets and (c,d) muon+jets events with (a,c) exactly 
	one and (b,d) at least two $b$-tagged jets. 
	The hatched bands show the normalisation uncertainty in the signal and background contributions and the signal model systematic uncertainties. The first 
	and last bins contain underflow and overflow events, respectively.}
\label{fig:control_plots_LL}
\end{center}
\end{figure}

\section{Template fit}
\label{sec:fitting}

The decay width of the top quark is measured using a simultaneous template fit to distributions of two observables associated with the hadronic and 
semileptonic decay branches of $t\bar{t}$ events in the eight mutually exclusive analysis regions.
The observables are $m_{\ell b}$, which is the reconstructed invariant mass of the $b$-jet of the semileptonically decaying top quark and the corresponding 
lepton, and \bsdR, which is the angular distance between the $b$-jet $j_b$ and the closest light jet $j_l$, both originating from the hadronically decaying top quark.
The choice of \mlb is due to its good sensitivity to $\Gamma_t$ while 
being less sensitive to jet-related uncertainties compared to 
reconstructed masses of the hadronic decay branch. Despite the much lower 
sensitivity of \bsdR to $\Gamma_t$, it is beneficial to use it in the fit 
because it adds information from the hadronic top-quark decay branch and 
reduces leading jet-related and signal model 
systematic uncertainties in the combination with \mlb. 
Several other observables defined using the invariant
mass of, or angles between, the \ttbar decay products were tested
but were found to be less suitable because of larger jet-related or signal
model uncertainties.

Signal templates are generated by reweighting events at parton-level to Breit--Wigner distributions with alternative 
top-quark decay-width hypotheses.
A total of 54 templates for different values of $\Gamma_t$ are created: 50 templates cover the  
range $0.1 < \Gamma_t < 5.0$~GeV in steps of $\Delta \Gamma = 0.1$~GeV. Four additional templates are created 
for $\Gamma_t = 0.01,~6,~7,~8$ GeV to take into account very small and very large width values. 
The top-quark decay width in the nominal MC signal sample is $\Gamma_t = 1.33$~GeV corresponding to the 
NLO calculation. The reweighting method was validated using a signal MC sample generated with $\Gamma_t = 3.0$~GeV 
by comparing 
top-quark mass distributions of this sample with top-quark mass distributions obtained from the 
reweighting procedure at parton level, and no significant differences were observed.  The impact on the 
template distributions by varying
the decay width in the range of 0.7 to 3.0~GeV is shown in Figure~\ref{fig:width_templates}.

\begin{figure}[!htbp]
\begin{center} 
	\subfloat[]{\includegraphics[width=75mm]{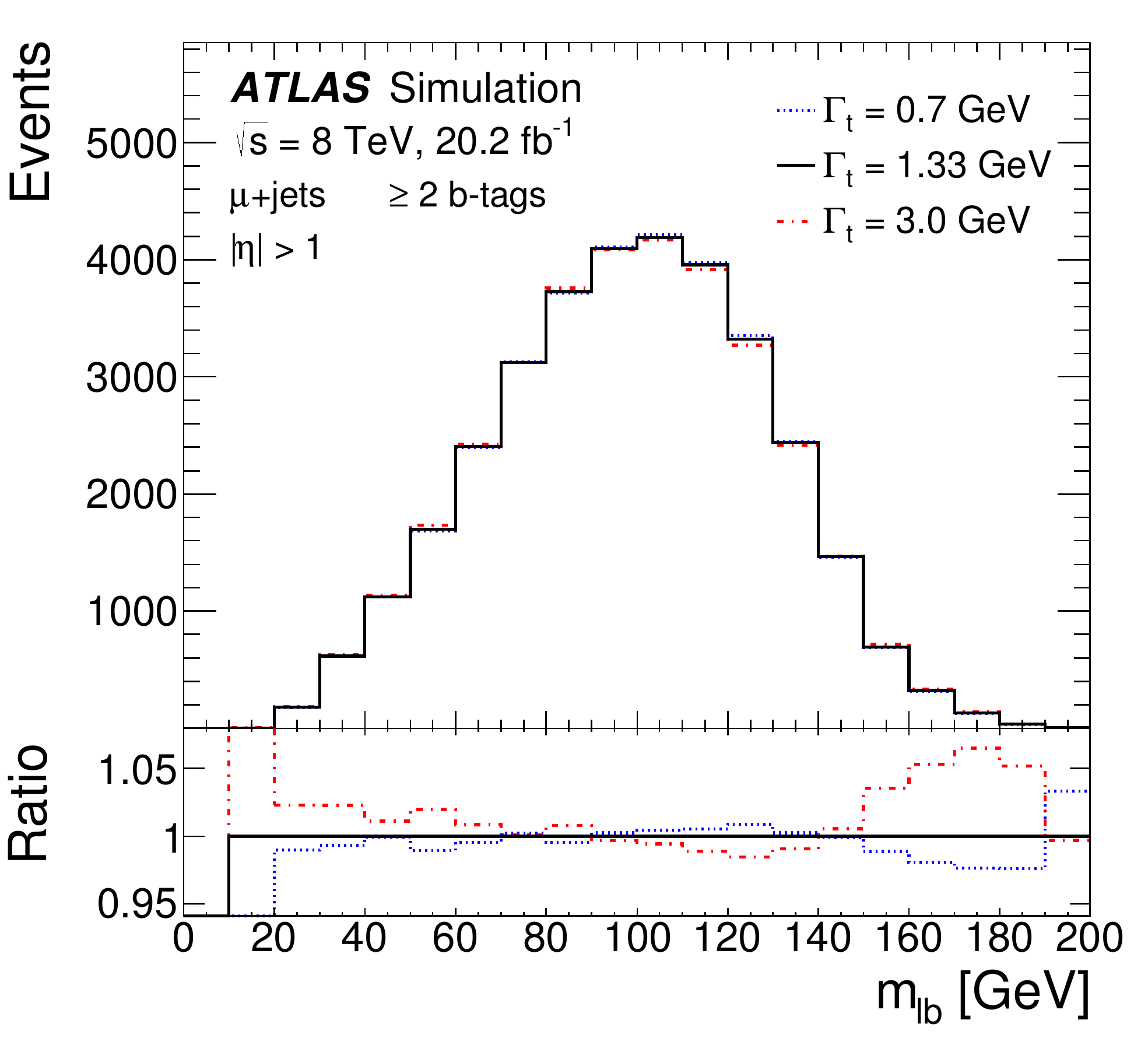} }
        \subfloat[]{\includegraphics[width=75mm]{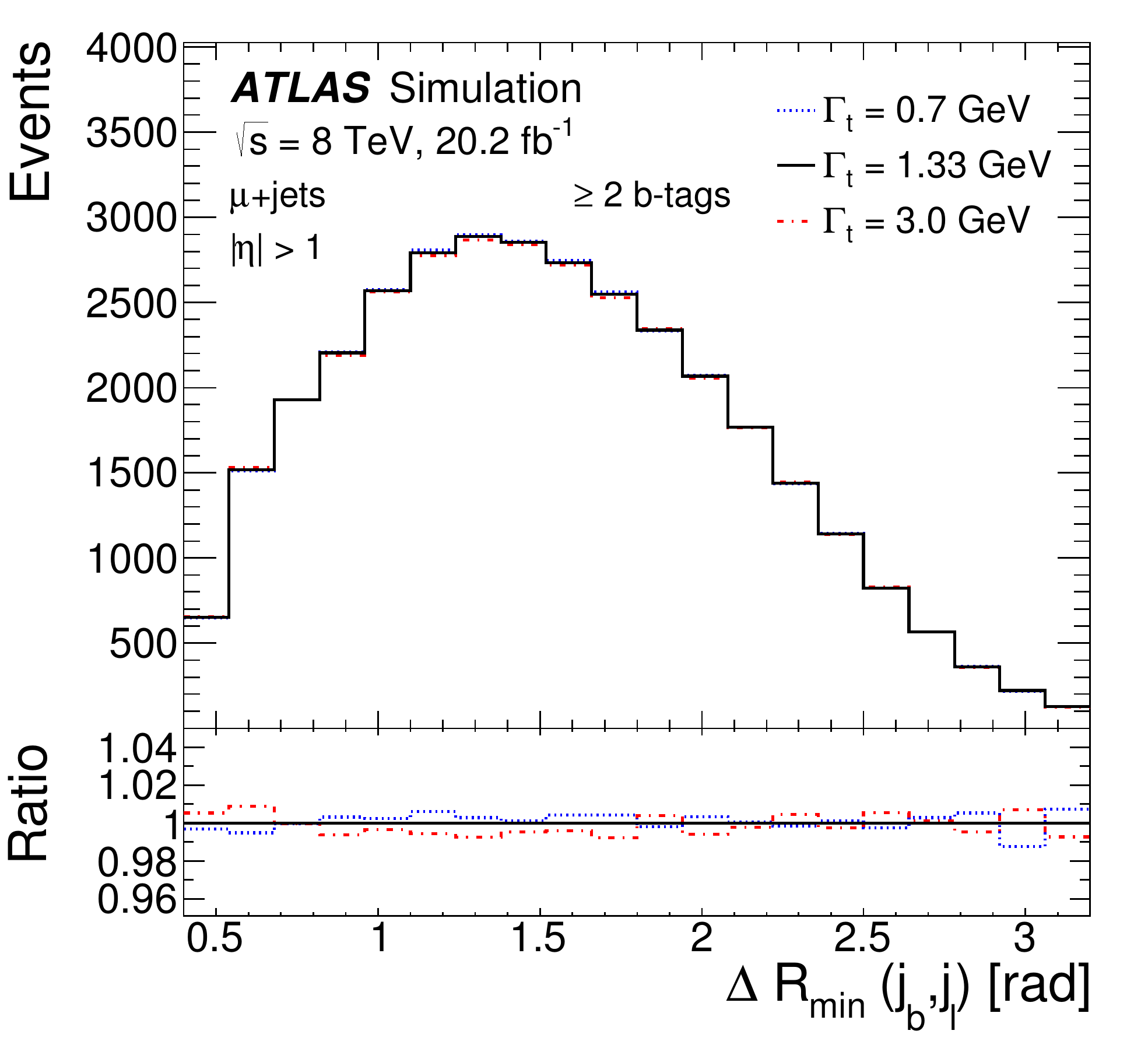} }
	\caption{Templates for (a) the reconstructed invariant mass of the $b$-jet of the semileptonically decaying top 
	quark and the corresponding lepton, \mlb, and (b) \bsdR, the angular distance between the $b$-jet $j_b$ associated with the hadronic top quark 
	and the closest light jet $j_l$ from the hadronically decaying $W$ boson, in the range $0.7 \leq \Gamma_t \leq 3.0$~GeV in the  muon+jets channel for events with at least two~$b$-tags in the $|\eta| > 1$ region. The lower panel shows the ratio of the templates with varied $\Gamma_t$ to the nominal template generated for a decay width of $\Gamma_t = 1.33$~GeV.}
\label{fig:width_templates}
\end{center}
\end{figure}

The binned likelihood fit to data uses these signal templates for the \ttbar contribution. Templates for all other 
processes, including single-top-quark production, are fixed. The effect on the result of using a fixed decay width in the single-top-quark template was found to be negligible.
The number of expected events per bin $i$ is given by

\begin{align*}
n_{i}= n_{\textrm{signal},i} + \sum\limits_{j=1}^{\textrm{B}} {n_{\mathrm{bkg},ji}}\,,
\end{align*}
where the index $j$ runs over all backgrounds. 
The likelihood for an observable $\mathcal{O}$ is defined as follows:

\begin{align}
 \mathscr{L} (\mathcal{O}|\Gamma_t ) = \prod\limits_{i=1}^{N_\textrm{bins}}  
 \mathrm{Poisson} (n_{\mathrm{data},i} | n_{i}(\Gamma_t) ) \cdot 
 \prod\limits_{j=1}^{\textrm{B}}\frac{1}{\sqrt{2\pi}\sigma_{\mathrm{bkg},j}}\exp \left( {\frac{-(n_{\mathrm{bkg},j}-\hat n_{\mathrm{bkg},j})^2}{2\sigma ^{2}_{\mathrm{bkg},j}}} \right)\,,
 \label{eq:likelihood}
\end{align}

where $N_{\mathrm{bins}}$ is the number of bins in a template and  
$n_{\mathrm{data},i}$ is the number of data events in each bin $i$. 
The number of events from a background source $j$, $n_{\mathrm{bkg},j}$, is 
obtained from $n_{\mathrm{bkg},ji}$ by summing over all bins $i$. 
This number of background events varies in the fit but it is constrained by Gaussian terms 
where $\hat n_{\mathrm{bkg},j}$ is 
the expected number of background events for source $j$ and $\sigma_{\mathrm{bkg},j}$ is its 
uncertainty. 
The total number of signal events is a free parameter of the fit. For each background source $j$ only one fit parameter $n_{\mathrm{bkg},j}$ is used for all $b$-tag bins, lepton channels and $|\eta|$ regions except for the multijet 
background. For the latter, separate parameters are 
defined for each analysis region. 

The uncertainties used as constraints in Eq.~(\ref{eq:likelihood}) on the $W$+jets background components 
normalisation originating from data-driven calibration (see Section~\ref{sec:samples}) amount to 7\% for $W+b\bar{b}$ and 
$W+c\bar{c}$, 25\% for $W+c$, and 5\% for $W$+light jets events.  
The uncertainty in the multijet background is taken from the matrix method and 
amounts to 30\%. For the $Z$+jets and diboson events, a 4\% theory uncertainty in the inclusive cross-section is 
applied together with a 24\% uncertainty per additional jet added in quadrature, which covers 
the extrapolation 
to higher jet multiplicities based on MC studies, resulting in an uncertainty of 48\% for events with four jets.
The uncertainty in single-top-quark production amounts to 17\% and considers the variation of initial- and final-state radiation in the $t$-channel MC samples and accounts for extra jets in single-top-quark events.

The fit is performed for 55 templates (54 obtained from the reweighting algorithm and the nominal one). The combined likelihood, defined 
as the product of two Poisson terms as given in Eq.~(\ref{eq:likelihood}), one for each observable, multiplied by the Gaussian constraints, 
is maximised for every value of $\Gamma_t$. 
The measured top-quark decay width is extracted from the minimum of a quadratic fit to the negative logarithm of the likelihood values.
The fit method was validated using pseudo-experiments, and the correlation 
between the two observables was examined. In each pseudo-experiment the content of the bins 
of the \mlb and \bsdR distributions 
are varied according to the Poisson and Gaussian distributions to take into account 
the expected number of events per bin and the background constraints, respectively.   
These pseudo-experiments 
are used for a linearity test and to produce pull distributions. The pull is 
defined as the difference between the fitted value $\Gamma_t$ and the input 
value divided by the estimated uncertainty of the fit result. 
No deviations from the expectation were found for various decay width values 
within $1.1 < \Gamma_t < 4.0$~GeV. For smaller decay width values the pull width decreases 
since the $\Gamma_t$ distribution approaches a limit of 0~GeV. However, this does 
not affect the result and the fit method is stable and unbiased. Additional pseudo-experiments revealed that the 
small correlation between $m_{\ell b}$ and \bsdR of about (0.1--2.8)\% does not affect the fit result. The observables 
are thus treated as independent.  

\section{Systematic uncertainties}
\label{sec:syst}

Systematic uncertainties affect the normalisation of signal and background 
and the shape of the distributions sensitive to the top-quark decay width. 
Individual sources of systematic uncertainty are considered uncorrelated 
and are summed in quadrature to determine the total uncertainty. Correlations 
of systematic uncertainties from the same source are fully maintained for all analysis regions.
Pseudo-experiments are used to estimate the impact of the different sources 
of uncertainty according to the following procedure. 
For each source of systematic uncertainty, templates
corresponding to the respective up and down variations are created.
These variations consider shape and acceptance changes from the systematic uncertainty
source under study. Pseudo-data sets are generated by imposing Poisson 
fluctuations and Gaussian fluctuations on the background contributions (to account for the Gaussian constraints) to 
each bin, as described in Section~\ref{sec:fitting}. Then the nominal and 
varied templates are used to perform a fit to pseudo-data. 
For each systematic variation 2000 of these pseudo-experiments were 
performed, and the differences between the means of the fitted-value distribution 
using the nominal templates 
and the up and down variations are quoted as the systematic 
uncertainty from this source. The systematic uncertainties in the 
measurement of the top-quark decay 
width are summarised in Table~\ref{tab:SystFull}.

\subsection{Uncertainties in detector modelling}

The systematic uncertainties arising from charged leptons are classified 
into several categories, related to the reconstruction and 
trigger efficiency, the identification and the lepton momentum scale and 
resolution. This leads to five (six) components of uncertainties associated 
with the electron (muon). 

Jet-related uncertainties arise from the jet reconstruction 
efficiency, the jet vertex fraction requirement, the jet energy 
resolution (JER) and the jet energy scale. The JES and its 
uncertainties were derived by combining information from test-beam data, LHC collision 
data and simulation~\cite{ATLASJetEnergyMeasurement,ATLAS-CONF-2015-037}. The JES calibration is described in Section~\ref{sec:objects_reco}. 
The jet energy scale uncertainty is split into 26 \pt- and $\eta$-dependent 
sources, treated independently. It is the largest of the detector modelling uncertainties 
in this analysis. 

The JER was evaluated separately for data and simulation using two in situ 
techniques~\cite{ATLASJetEnergyMeasurement}, improved by additional in situ measurements 
using dijet, photon+jet or $Z$+jet processes. For low-\pt jets, a significant contribution 
to the JER uncertainty comes from pile-up, measured as in Ref.~\cite{ATLAS-CONF-2015-037}. 
The JER uncertainty consists of 11 components and represents an important uncertainty 
in this measurement. The systematic uncertainties originating from these components are summed in quadrature. The symmetrised difference is the quoted systematic uncertainty in the JER.

The per-jet efficiency to pass the JVF selection is evaluated in $Z(\to \ell^+\ell^-)$+1-jet 
events in data and simulation~\cite{Aad:2015ina}. Motivated by this study, the uncertainty is 
estimated by changing
the JVF requirement value, increasing and decreasing it by 0.1, and repeating
the analysis using this modified value.

Energy scale and resolution correction uncertainties of both the leptons and jets are 
propagated into the calculation of \met.
Contributions from energy deposits not associated with any jet and due to soft-jets (7 GeV~$< p_\text{T} <$~20 GeV) are also considered and treated as fully
correlated with each other.
A further \met uncertainty accounts for mis-modeling of pileup energy deposits. 

The jet-flavour-dependent efficiencies of the $b$-tagging algorithm are 
calibrated using data. The $b$-tagging efficiency is corrected to match the efficiency measured in the \ttbar data events using the probability density function
calibration method~\cite{ATLAS-CONF-2014-004,ATLAS-CONF-2014-046} based on a combinatorial likelihood applied to a data sample of dileptonic \ttbar events. The mistag rate for $c$-jets is measured using $D^*$ mesons, the one for light jets is measured using jets with impact parameters and secondary vertices consistent with a negative lifetime~\cite{Aad:2015ydr,ATLAS-CONF-2014-046}.
Efficiencies for $b$- and $c$-jets are corrected in 
simulations by $p_\text{T}$-dependent scale factors. 
For light jets, these scale factors also depend on the pseudorapidity. Six independent 
sources of uncertainty affecting the $b$-tagging efficiency and four affecting the $c$-tagging 
efficiency are considered~\cite{ATLAS-CONF-2014-004}. For the mistagging of light-quark jets, 12 uncertainties which depend on jet \pt and $\eta$~\cite{ATLAS-CONF-2014-046} are considered.

\subsection{Uncertainties in background modelling}

The uncertainties in the background normalisation are included as Gaussian 
constraints in the fit (see Eq.~(\ref{eq:likelihood})) and thus contribute to the overall statistical uncertainty.     

To estimate the uncertainty in the shape modelling of the $W$+jets background, each of   
its flavour components ($W$+$b\bar{b}/c\bar{c}$, $W$+$c$ and $W$+light) is allowed to vary 
independently in the fit within its uncertainty, corresponding to the uncertainty in the 
calibration factors. The shape uncertainty of the $W$+jets contribution is retrieved by varying 
one component while fixing the other two to their respective normalisations, as given in Section~\ref{sec:samples}.

Two simulated samples are compared to estimate the modelling uncertainty of single-top events. 
The baseline MC event generator for $Wt$ production uses the diagram removal 
technique~\cite{Frixione:2008yi} to account for the overlap with $t\bar{t}$ events. This sample is 
compared to a sample generated with the inclusive diagram subtraction 
technique~\cite{Frixione:2008yi}. The difference is then symmetrised, i.e.~the difference of 
the two-point comparison is taken as the uncertainty on both sides of the nominal result.

For the multijet background an uncertainty in the total yield of $\pm 30$\%
is assigned. Furthermore, two shape uncertainties are defined by
varying the control samples used to obtain the efficiencies used in
the matrix method to relate the two identification levels for objects
considered as fake or non-prompt and prompt leptons, respectively.

The background yields estimated from MC simulation are affected by the luminosity 
uncertainty of
1.9\%~\cite{lumi_8TeV}, which is propagated to the constraints 
on the background yields.

\subsection{Uncertainties in signal modelling}

Several uncertainties affect the shape of the \ttbar signal contributions. 
The uncertainties due to 
initial- and final-state radiation are determined using two dedicated \textsc{Powheg+Pythia} 
samples (see Section~\ref{sec:samples}) generated with varied parameter values giving more or less radiation. 
Pseudo-data is created using each sample, and the largest variation of the top-quark 
decay width 
from the nominal is taken as an uncertainty and then symmetrised.  

The \textsc{Powheg} MC event generator interfaced to \textsc{Pythia} is compared to \textsc{ Powheg} interfaced 
with \textsc{Herwig} to 
estimate the uncertainty due to the parton shower and the hadronisation model. To estimate the 
uncertainty in the choice of the \ttbar event generator, the full difference between \textsc{ Powheg} 
and MC$@$NLO event generators, both interfaced with \textsc{Herwig}, is evaluated. This is the largest
signal modelling uncertainty in this measurement. The uncertainty 
of the colour reconnection modelling is estimated by comparing the nominal \ttbar sample to a \textsc{Powheg} sample interfaced with {\textsc Pythia} with the Perugia parameter tune ``P2012IoCR``~\cite{Skands:2010ak} for colour reconnection. This tune has a slightly lower colour reconnection strength than the default tune, which affects the corresponding colour strings, and is combined with a slightly modified colour reconnection algorithm.
The uncertainty in the underlying-event modelling is determined by 
comparing the nominal \ttbar sample with a \textsc{Powheg} sample interfaced with \textsc {Pythia} employing the Perugia parameter tune “P2012mpiHI''~\cite{Skands:2010ak} for multiparton interactions. This tune increases the number of multi-parton interaction (MPI) scatterings, which leads to an increase in MPI minijets. This is realised by a larger $\alpha_S$ value associated with the MPI.
The uncertainties due to these four sources are taken as the difference between the nominal and the varied sample and symmetrised, i.e. the full difference is taken as the positive and negative uncertainty.

Following the PDF4LHC~\cite{ref:pdf4lhc} recommendations, three different PDF sets are compared using a reweighting technique for the signal $t\bar{t}$ MC sample to estimate the uncertainty due to the PDF set choice: 
CT10 NLO (nominal PDF set)~\cite{ct101}, MSTW 2008 68\% CL NLO~\cite{mstw1} and NNPDF 2.3 NLO~\cite{nnpdf}.
Each PDF set has a different prescription for using its error sets to evaluate the uncertainty: the CT10 set uses a symmetric Hessian matrix, 
the MSTW set uses an asymmetric Hessian matrix and the NNPDF set uses a standard deviation for the uncertainty calculation. For the three PDF sets, the variations for all different PDF parameters are evaluated within the corresponding set. Half of the width of the largest deviation from nominal among all three sets is taken as the PDF uncertainty.

\subsection{NLO and off-shell effects in the top-quark decay}
The \ttbar MC simulation utilised to extract the decay width 
uses NLO matrix elements for top-quark pair production and LO matrix elements with approximate implementation of finite-width and interference effects for the decay of the top quarks.
A theoretical study~\cite{powres} performed in the $e\mu$ decay channel of the
\ttbar system indicates that taking into account off-shell effects, which include 
the contributions from \ttbar and $Wt$ single-top production 
as well as their interference,
is important for the precision measurements of top-quark properties.  
However, there is no MC implementation yet of NLO decay and off-shell effects 
for the lepton+jets final state. 
The potential impact of ignoring these effects was tested in two different ways. First, the 
parton-level \mlb distribution of a sum of \ttbar and $Wt$ single-top 
contributions without these effects taken into account was reweighted to 
the \mlb distribution provided by the authors of Ref.~\cite{powres} which 
corresponds to the  $WWb\bar{b}$ calculation at NLO. 
Second, the measurement was repeated 
requiring $\mlb < 150$~\GeV, limiting the analysis to the region where these effects are expected to be suppressed according to 
Ref.~\cite{powres}. Both cross-checks
yield a difference in the measured top-quark decay width 
of less than 0.5~GeV. 
For more precise future measurements, it would be beneficial to have an MC simulation 
providing an NLO description of the top-quark decay accounting for off-shell effects.

\begin{table}[t]
\begin{small}
\begin{center}
\caption{Summary of systematic uncertainties in the top-quark decay width measurement.} 
\renewcommand{\arraystretch}{1.25}
\begin{tabular}{|l|c|}
\hline
Source & Uncertainty [GeV]\\
 \hline\hline
\textbf{Detector model} &  \\
Electron &  $^{+0.14}_{-0.07}$  \\
Muon     &  $^{+0.11}_{-0.06}$  \\ 
Jet energy scale &  $^{+0.42}_{-0.30}$ \\ 
Jet energy resolution &  $\pm 0.27$          \\ 
Jet vertex fraction &  $^{+0.13}_{-0.03}$  \\ 
Jet reconstruction efficiency & $\pm 0.03$ \\          
Missing transverse momentum & $\pm 0.01$ \\
$b$-Tagging &  $^{+0.32}_{-0.24}$  \\ \hline
 \textbf{Signal model} & \\  
 ME event generator & $\pm 0.41$ \\
 Colour reconnection &  $\pm 0.19$ \\
 Underlying event & $\pm 0.11$ \\
 Radiation & $\pm 0.07$ \\
 PDF & $\pm 0.06$ \\
 PS/hadronisation & $\pm 0.05$ \\
 \hline
\textbf{Background model} &  \\
Multijet &  $^{+0.04}_{-0.00}$ \\
 $W$+jets & $\pm 0.02$ \\
 Single top & $< 0.01$\, \\
 \hline
 Template statistical uncertainty &  $\pm 0.07$ \\
\hline 
Luminosity    &  $^{+0.03}_{-0.00}$  \\ 
 \hline\hline
Total systematic uncertainty    &  $^{+0.79}_{-0.68}$  \\ 
 \hline
\end{tabular}

\label{tab:SystFull}

\end{center}
\end{small}
\end{table}

\subsection{Template statistical uncertainty}

To estimate the systematic uncertainty arising from the limited MC sample size
used to produce the templates, the content of each bin of the signal and background templates 
is varied within its statistical uncertainty and a fit to the nominal 
distribution is repeated. The MC statistical uncertainty is derived with and without taking into account the correlations between the templates and both estimates yield consistent results. 
The standard deviation of the distribution of 
top-quark decay width values obtained from the fits with the fluctuated 
templates is quoted as the systematic effect from the template statistical uncertainty.

\section{Result}
\label{sec:result}

The binned likelihood template fit is applied to the data using the 
concatenated distributions of $m_{\ell b}$ and \bsdR in the eight 
analysis regions. 
Figure~\ref{fig:post_fit} shows post-fit distributions for $m_{\ell b}$ and \bsdR. 
The post-fit yields of the $t\bar{t}$ signal and each background contribution 
are summarised in Table~\ref{tab:post_fit}. 

\begin{figure}[!htbp]
\begin{center}
	\subfloat[]{
         \includegraphics[width=145mm]{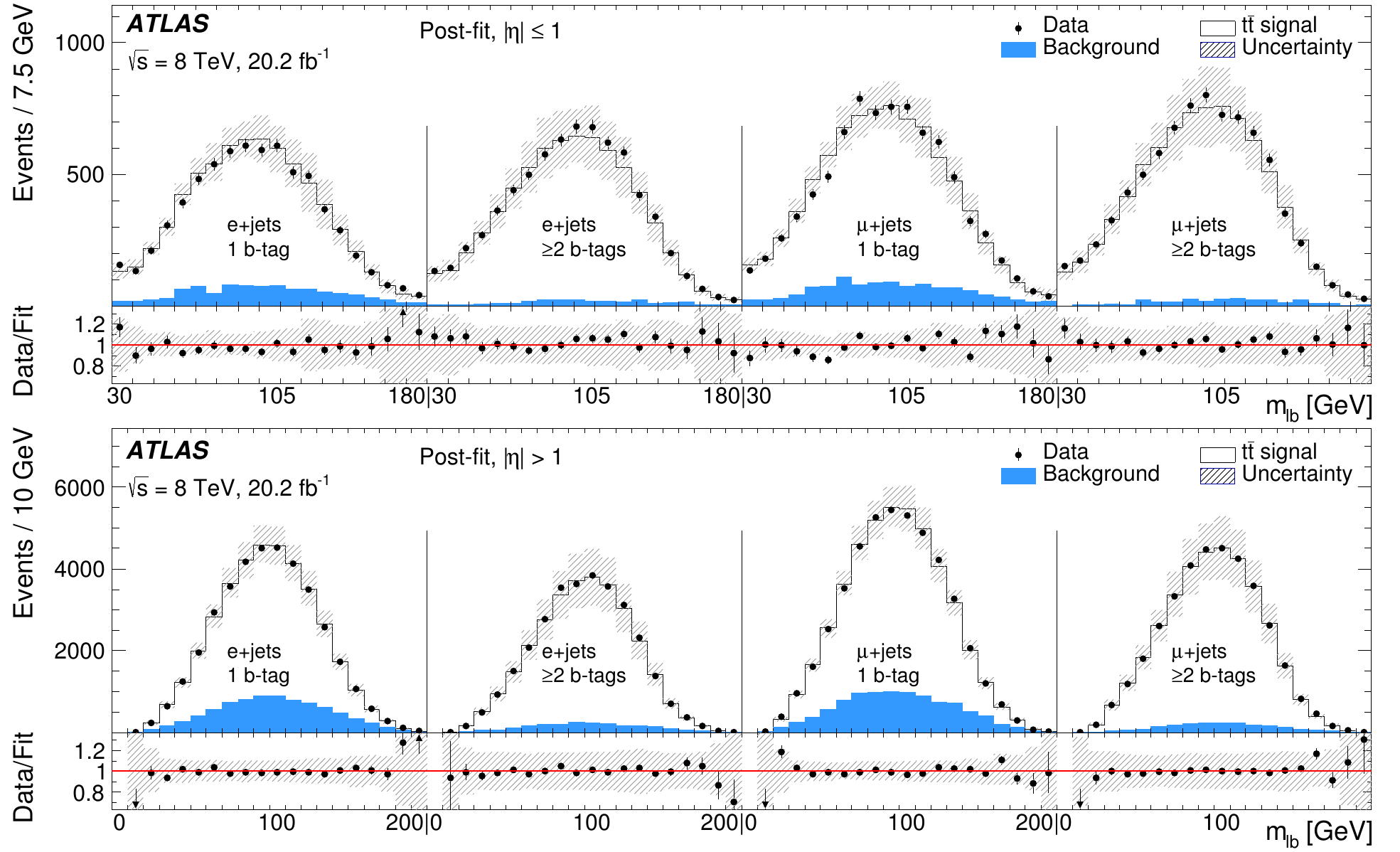} }\\
	\subfloat[]{
         \includegraphics[width=145mm]{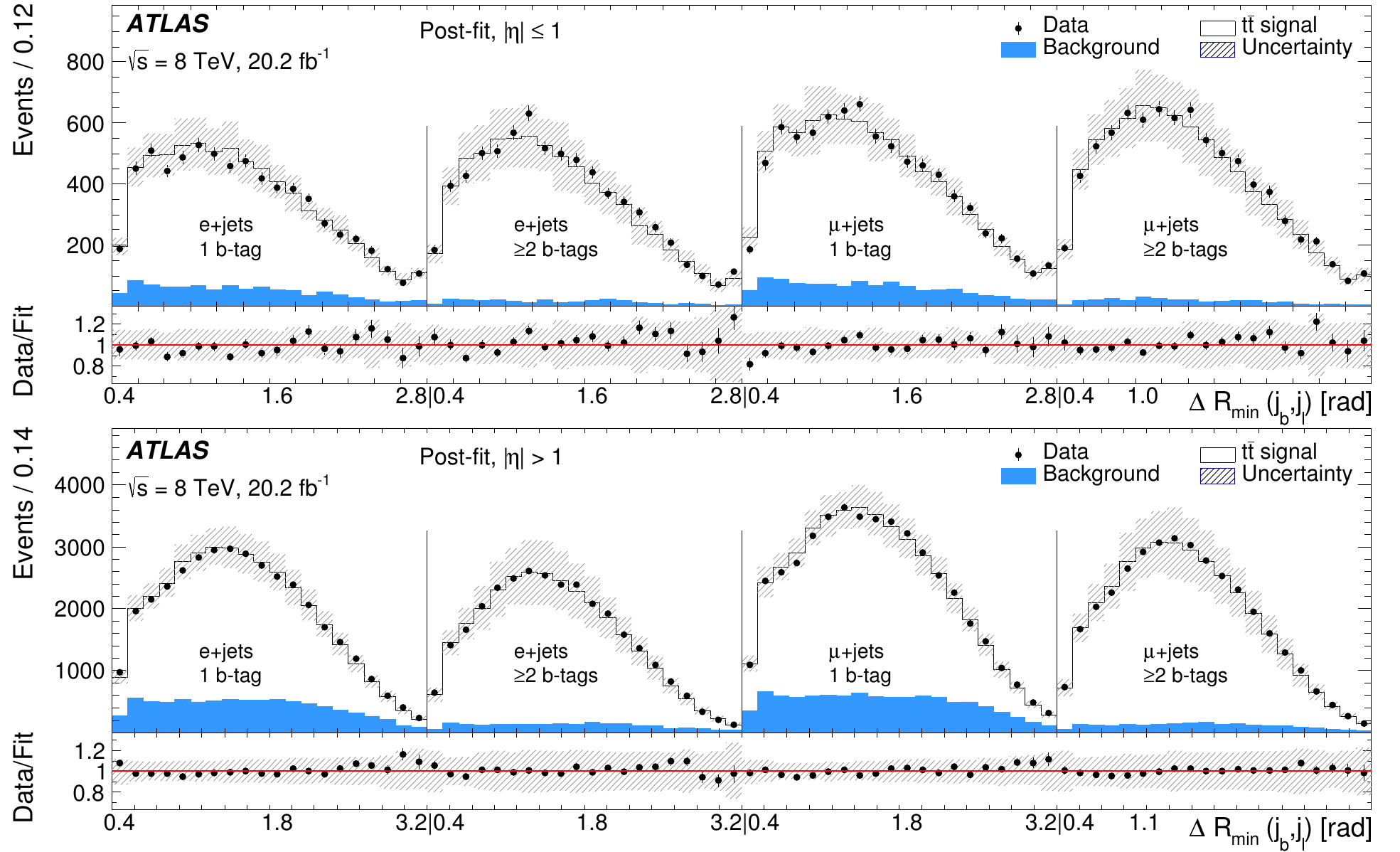} }
         \caption{Post-fit distributions based on the best-fit templates for (a) $m_{\ell b}$ and (b) \bsdR. The background contributions 
	 are combined. 
	 The lower panel shows the ratio of data to post-fit sum of $t\bar{t}$ signal and background. The eight 
	 analysis regions 
	 corresponding to different $b$-tag multiplicity and jet pseudorapidity are shown. 
	The vertical lines show the boundaries between the binned variables in different lepton and $b$-tag regions.
	The hatched band shows the total uncertainty. The systematic uncertainties are calculated bin-by-bin from the systematic variations by adding differences in quadrature. Then, statistical and systematic uncertainties are added in quadrature to obtain the quoted total uncertainty.}
		\label{fig:post_fit}
\end{center}
\end{figure}

\begin{table}[t]
\centering
\begin{small}
\captionof{table}{Post-fit yields of the $t\bar{t}$ signal and background contributions. The yields 
represent the sum of the number of events in each of the eight analysis regions. Only the normalisation 
uncertainties are shown.} 
\begin{minipage}{\linewidth}\centering
\begin{tabular}{|l|r|}
\hline
Sample & Post-fit yields \\
\hline
$t\bar{t}$ & 156360 $\pm$ ~~750 \\
Single top & 5700 $\pm$ ~~930 \\
$W$ + bb/cc & 7060 $\pm$ ~~510 \\
$W$ + c & 1650 $\pm$ ~~550 \\
$W$ + light & 1603 $\pm$ ~~~~65  \\
$Z$ + jets & 2770 $\pm$ ~~710  \\
Diboson & 320 $\pm$ ~~240  \\
Multijet & 6070 $\pm$ ~~380 \\
\hline
Total & 181600 $\pm$ 1700  \\
\hline
Data & 181536 \hspace{10.2mm} \\
\hline
\end{tabular}

\label{tab:post_fit}
\end{minipage}
\end{small}
\end{table}

The likelihood curve obtained from the fit can be seen in Figure~\ref{fig:likelihoodScan} together with a 
quadratic fit to the likelihood points. 
The statistical uncertainty, which includes contributions from the data statistics and the uncertainties in 
the backgrounds normalisation, is extracted from the likelihood curve's 
width at $-2\Delta \ln(\mathscr{L})=1$ around the minimum. 
The likelihood values are shifted so that the minimum corresponds to  $-2\Delta\ln(\mathscr{L}) = 0$.
\begin{figure}[!hbpt]
\begin{center} 
	\includegraphics[width=90mm]{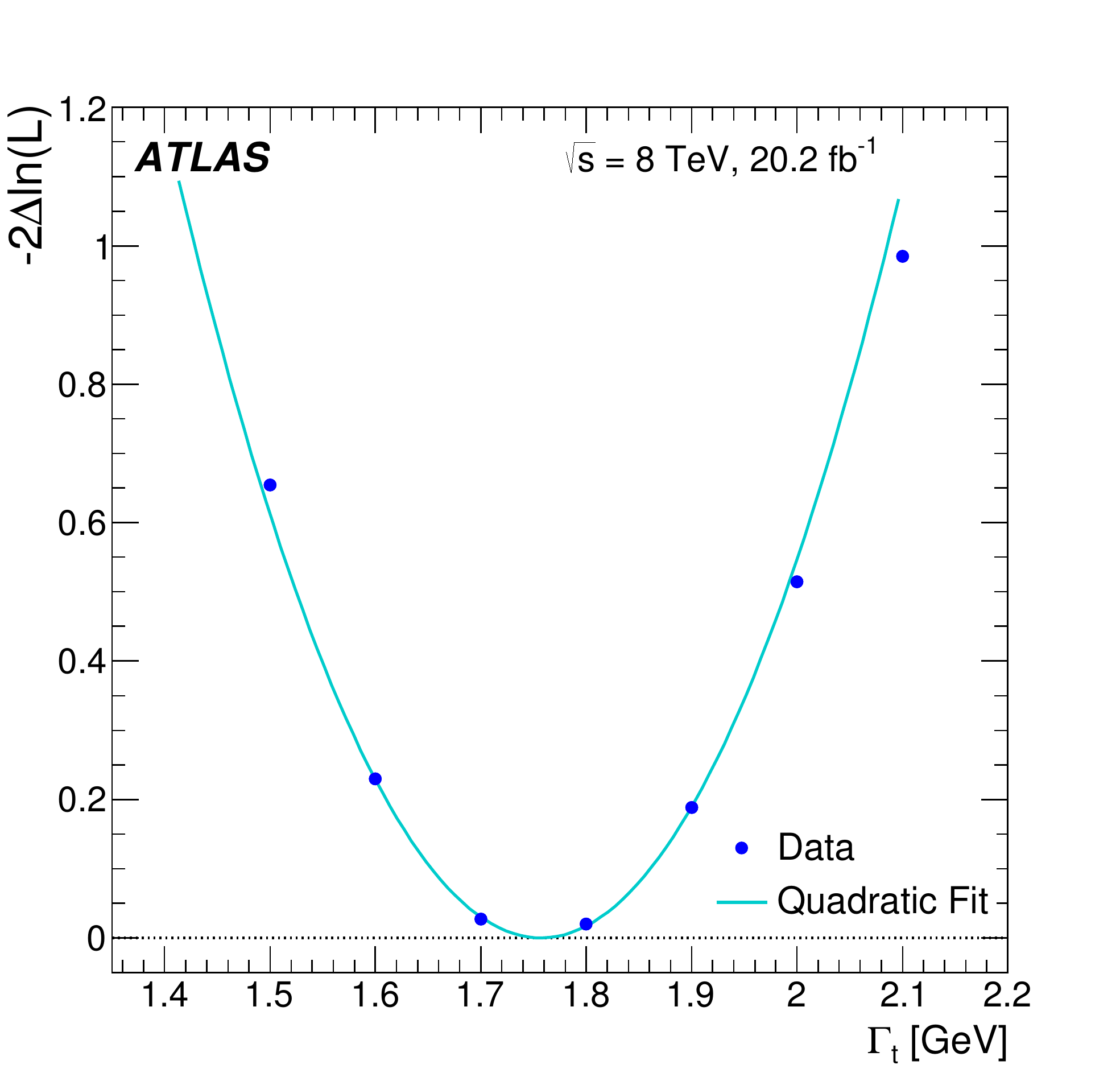}
	\caption{Twice the negative logarithm of the likelihood obtained from the binned likelihood template fit to data. 
	A quadratic fit is performed around the minimum.} 
\label{fig:likelihoodScan}
\end{center}
\end{figure}

The measured decay width for a top-quark mass of 172.5~GeV is 

\begin{align*}
 \Gamma_t = 1.76 \pm 0.33~(\text{stat.})~ ^{+0.79}_{-0.68}~(\text{syst.})~\text{GeV} = 1.76^{+0.86}_{-0.76}~\text{GeV}\,, 
\end{align*}

in good agreement with the SM prediction of 1.322~GeV~\cite{Gao:2012ja}. 
A consistency check was performed by repeating the measurement 
in the individual $b$-tag regions and confirms that the results are consistent with 
the measured value. A fit based only on the observable \mlb leads to a total uncertainty which is about 0.3~GeV larger.

In comparison to the previous direct top-quark decay width measurement in Ref.~\cite{Aaltonen:2013kna}, the total uncertainty 
of this measurement is smaller by a factor of around two. However, this result is still less precise than indirect 
measurements and, thus, 
alternative (BSM) models discussed in Section~\ref{sec:intro} cannot be ruled out with the current sensitivity.

The impact of the assumed top-quark mass on the decay width measurement is estimated by varying the mass around the nominal value of $m_t=172.5$~GeV. Changing the top-quark mass by $\pm 0.5$~GeV leads to a shift in the measured top-quark decay width of 
up to around 0.2~GeV.

\FloatBarrier
\section{Conclusion}
\label{sec:conclusion}

A direct measurement of the decay width of the top quark 
exploiting \ttbar events in the lepton+jets channel was performed using 
data taken in proton--proton collisions at $\sqrt{s} =  8$~TeV corresponding to an integrated luminosity of 20.2~fb$^{-1}$ recorded by the ATLAS detector at the LHC.
The decay width of the top quark is extracted using a binned likelihood 
template fit to data based on two observables related to the hadronic and the 
semileptonic decay branches of the \ttbar pair. The top-quark decay width is 
measured to be 
$\Gamma_t = 1.76 \pm 0.33~(\text{stat.})~ ^{+0.79}_{-0.68}~(\text{syst.})~\text{GeV}$ 
for $m_t = 172.5$~GeV, which is in a good agreement with SM predictions.

\section*{Acknowledgements}


We thank CERN for the very successful operation of the LHC, as well as the
support staff from our institutions without whom ATLAS could not be
operated efficiently.

We acknowledge the support of ANPCyT, Argentina; YerPhI, Armenia; ARC, Australia; BMWFW and FWF, Austria; ANAS, Azerbaijan; SSTC, Belarus; CNPq and FAPESP, Brazil; NSERC, NRC and CFI, Canada; CERN; CONICYT, Chile; CAS, MOST and NSFC, China; COLCIENCIAS, Colombia; MSMT CR, MPO CR and VSC CR, Czech Republic; DNRF and DNSRC, Denmark; IN2P3-CNRS, CEA-DRF/IRFU, France; SRNSFG, Georgia; BMBF, HGF, and MPG, Germany; GSRT, Greece; RGC, Hong Kong SAR, China; ISF, I-CORE and Benoziyo Center, Israel; INFN, Italy; MEXT and JSPS, Japan; CNRST, Morocco; NWO, Netherlands; RCN, Norway; MNiSW and NCN, Poland; FCT, Portugal; MNE/IFA, Romania; MES of Russia and NRC KI, Russian Federation; JINR; MESTD, Serbia; MSSR, Slovakia; ARRS and MIZ\v{S}, Slovenia; DST/NRF, South Africa; MINECO, Spain; SRC and Wallenberg Foundation, Sweden; SERI, SNSF and Cantons of Bern and Geneva, Switzerland; MOST, Taiwan; TAEK, Turkey; STFC, United Kingdom; DOE and NSF, United States of America. In addition, individual groups and members have received support from BCKDF, the Canada Council, CANARIE, CRC, Compute Canada, FQRNT, and the Ontario Innovation Trust, Canada; EPLANET, ERC, ERDF, FP7, Horizon 2020 and Marie Sk{\l}odowska-Curie Actions, European Union; Investissements d'Avenir Labex and Idex, ANR, R{\'e}gion Auvergne and Fondation Partager le Savoir, France; DFG and AvH Foundation, Germany; Herakleitos, Thales and Aristeia programmes co-financed by EU-ESF and the Greek NSRF; BSF, GIF and Minerva, Israel; BRF, Norway; CERCA Programme Generalitat de Catalunya, Generalitat Valenciana, Spain; the Royal Society and Leverhulme Trust, United Kingdom.

The crucial computing support from all WLCG partners is acknowledged gratefully, in particular from CERN, the ATLAS Tier-1 facilities at TRIUMF (Canada), NDGF (Denmark, Norway, Sweden), CC-IN2P3 (France), KIT/GridKA (Germany), INFN-CNAF (Italy), NL-T1 (Netherlands), PIC (Spain), ASGC (Taiwan), RAL (UK) and BNL (USA), the Tier-2 facilities worldwide and large non-WLCG resource providers. Major contributors of computing resources are listed in Ref.~\cite{ATL-GEN-PUB-2016-002}.

\clearpage

\addcontentsline{toc}{part}{References}
\printbibliography

\newpage 
\begin{flushleft}
{\Large The ATLAS Collaboration}

\bigskip

M.~Aaboud$^\textrm{\scriptsize 137d}$,
G.~Aad$^\textrm{\scriptsize 88}$,
B.~Abbott$^\textrm{\scriptsize 115}$,
O.~Abdinov$^\textrm{\scriptsize 12}$$^{,*}$,
B.~Abeloos$^\textrm{\scriptsize 119}$,
S.H.~Abidi$^\textrm{\scriptsize 161}$,
O.S.~AbouZeid$^\textrm{\scriptsize 139}$,
N.L.~Abraham$^\textrm{\scriptsize 151}$,
H.~Abramowicz$^\textrm{\scriptsize 155}$,
H.~Abreu$^\textrm{\scriptsize 154}$
% ...
\bigskip
\\
% ...
$^{12}$ Institute of Physics, Azerbaijan Academy of Sciences, Baku, Azerbaijan\\
% ...
$^{88}$ CPPM, Aix-Marseille Universit{\'e} and CNRS/IN2P3, Marseille, France\\
% ...
$^{115}$ Homer L. Dodge Department of Physics and Astronomy, University of Oklahoma, Norman OK, United States of America\\
% ...
$^{119}$ LAL, Univ. Paris-Sud, CNRS/IN2P3, Universit{\'e} Paris-Saclay, Orsay, France\\
% ...
$^{137}$ $^{(a)}$ Facult{\'e} des Sciences Ain Chock, R{\'e}seau Universitaire de Physique des Hautes Energies - Universit{\'e} Hassan II, Casablanca; $^{(b)}$ Centre National de l'Energie des Sciences Techniques Nucleaires, Rabat; $^{(c)}$ Facult{\'e} des Sciences Semlalia, Universit{\'e} Cadi Ayyad, LPHEA-Marrakech; $^{(d)}$ Facult{\'e} des Sciences, Universit{\'e} Mohamed Premier and LPTPM, Oujda; $^{(e)}$ Facult{\'e} des sciences, Universit{\'e} Mohammed V, Rabat, Morocco\\
% ...
$^{139}$ Santa Cruz Institute for Particle Physics, University of California Santa Cruz, Santa Cruz CA, United States of America\\
% ...
$^{151}$ Department of Physics and Astronomy, University of Sussex, Brighton, United Kingdom\\
% ...
$^{154}$ Department of Physics, Technion: Israel Institute of Technology, Haifa, Israel\\
$^{155}$ Raymond and Beverly Sackler School of Physics and Astronomy, Tel Aviv University, Tel Aviv, Israel\\
% ...
$^{161}$ Department of Physics, University of Toronto, Toronto ON, Canada\\
% ...
$^{a}$ Also at Department of Physics, King's College London, London, United Kingdom\\
$^{b}$ Also at Institute of Physics, Azerbaijan Academy of Sciences, Baku, Azerbaijan\\
$^{c}$ Also at Novosibirsk State University, Novosibirsk, Russia\\
$^{d}$ Also at TRIUMF, Vancouver BC, Canada\\
$^{e}$ Also at Department of Physics {\&} Astronomy, University of Louisville, Louisville, KY, United States of America\\
% ...
$^{*}$ Deceased
\end{flushleft}


\end{document}